# Ideal Quantum Gases with Planck Scale Limitations

Rainer Collier

Institute of Theoretical Physics, Friedrich-Schiller-Universität Jena

Max-Wien-Platz 1, 07768 Jena, Germany

E-Mail: rainer@dr-collier.de

**Abstract.** A thermodynamic system of non-interacting quantum particles changes its statistical distribution formulas if there is a universal limitation for the size of quantum leaps $\Delta E \leq E_*$ ( $E_*$ Planck energy). By means of a restriction of the a priori equiprobability postulate one can arrive at a thermodynamic foundation of these corrected distribution formulas. The number of microstates $\Omega$ is determined by means of a suitable counting method and combined with thermodynamics via the Boltzmann principle $S = k_B \ln \Omega$. The result is that, for particle energies $\varepsilon_k \to E_*$, the thermodynamic difference between fermion and boson distribution vanishes. Both distributions then approximate a Boltzmann distribution. The wave-particle character of the quantum particles, too, can be influenced by choosing the size of the parameter values $k_B T / E_*$ and $\varepsilon_k / E_*$, as you can see from the associated fluctuation formulas. In the case of non-relativistic degeneration $T \to 0$, the critical quantities of Fermi momentum $p_F$ and Einstein temperature $T_E$ are corrected by the factor $q_0 = (1 - m_0 c^2 / E_*)$. For the Bose Einstein condensation there exists, in the condensation range $T < T_E$, a finite upper limit $N_0^{\max} = E_* / m_0 c^2$ for the number of particles in the ground state. In the relativistic high-temperature range $T \to \infty$, the energy densities of photon and neutrino radiation have finite limit values, which is of interest with regard to the start of cosmic expansion.

## 1 Introduction

It is a central proposition of almost all theories of quantum gravitation that there exists a universal lower bound $\Delta L \geq L_*$ for the uncertainty of length measurements, with $L_*$ being a length proportional to Planck's length. This is founded on the assumption that the geometry in spacetime ranges of the linear dimension $L_*$ has a structure that is grainy or, in any case, no longer continuous. Clearly, such an assumption violates the fundamental principles of quantum theory (e.g., Heisenberg's uncertainty principle, HUP) and of Einstein's relativity theory (e.g., the local Lorentz invariance). Recent years have seen a rapidly growing number of approaches that, aimed at a theory of quantum gravitation, attempt to overcome these contradictions and to install a minimum length scale that is independent of an observer. These approaches include loop quantum gravitation, string theory, noncommutative geometries, ansatzes for deformed Lorentz and Poincaré algebras, doubly special relativity theory (DSR) and various generalized uncertainty principles (GUP). Rather than discussing details here, we refer to some introductory articles [1] – [6] and special studies [7] – [15] on these topics. Not long ago, there appeared a comprehensive overview article on the physics of a possible minimum length scale and maximum momentum [40].

It is legitimate, though, to put the question whether such quantities as ´elementary length´ or ´energy limit´ can be arrived at also by an extension of the mathematical foundations of the theory of gravitation. In this case, such limits can be scalar parameters, which then appear in a continuous spacetime structure with local Lorentz symmetry (see, e.g., [39] and the literature cited there).

The assumption of the existence of a lower bound $L_*$ for length measurements is known to entail an upper bound $E_*$ for energy quanta $E$ with which spatial structures in Planck regions of size $L_*$ can still be resolved. There should be, e.g., no electromagnetic wave with a wavelength $\lambda < L_*$, i.e., no photon having an energy

$$E = \hbar\omega \sim \frac{c\hbar}{\lambda} > E_* = \hbar\Omega_* \sim \frac{c\hbar}{L_*} \ , \tag{1.1}$$

with which one could penetrate into spatial regions of the size $\lambda < L_*$ (see also [37]).

Therefore, let the further considerations be centred about the following hypothesis:

*As a consequence of a discrete microstructure of the spacetime manifold, quantum leaps between the energy eigenstates of a quantum system are universally limited to*

$$\Delta E \leq E_* \tag{1.2}$$

($E_* = M_* c^2 = \alpha M_{Pl} c^2$, $M_{PL} = \sqrt{\hbar c/G}$ ).

The quantity $E_*$ or $M_*$ needs to be determined experimentally or astrophysically. For the sake of simplicity, let us first use $\alpha = 1$, which makes $M_* = M_{Pl}$.

The present study is aimed at investigating which physical conclusions can be drawn from the above hypothesis (also cf. [16]). As a consequence of the hypothesis (1.2), the author already presented a corrected Planck radiation law in [17], which led to a modified Stefan-Boltzmann law for the internal energy $U(T)$ and, thus, for the energy density $u(T) = U(T)/V$ of the electromagnetic black-body radiation. The new caloric and thermal state equations read

$$U(T) = \frac{g_s \cdot V}{2\pi^2 (\hbar c)^3} \int_0^{E_*} \frac{\varepsilon^3 d\varepsilon}{\exp\left(\dfrac{\varepsilon}{k_B T}\right) - \left(1 - \dfrac{\varepsilon}{E_*}\right)} \ , \quad \varepsilon = \hbar\omega \ , \tag{1.3}$$

$$P(T) = -\frac{g_s \cdot (k_B T)}{2\pi^2 (\hbar c)^3} \int_0^{E_*} \frac{\varepsilon^2}{\left(1 - \frac{\varepsilon}{E_*}\right)} \ln\left[1 - \left(1 - \frac{\varepsilon}{E_*}\right) \exp\left(\frac{\varepsilon}{k_B T}\right)\right] d\varepsilon \ . \tag{1.4}$$

For $E_* \to \infty$, (1.3) and (1.4) are transformed back into the well-known Planck equations of state ($\sigma_B$ Stefan-Boltzmann constant, $g_s = 2$)

$$U(T) = \frac{4\sigma_B}{c} V T^4 \ , \quad \sigma_B = \frac{g_s \pi^5 k_B^4}{15 h^3 c^2} \ , \tag{1.5}$$

$$P(T) = \frac{1}{3} u(T) \ . \tag{1.6}$$

For $T \to \infty$, the energy density $u(T)$ associated with (1.3) has a finite limit

$$\lim_{T \to \infty} u(T) = \frac{g_s}{3} u_* \ , \tag{1.7}$$

with $u_*$ being the extremely high Planck energy density (see below). Therefore, the application of the state equations (1.3) and (1.4) to the early development phase of the universe was an interesting aspect of the hypothesis (1.2). It turned out that already a pure radiation cosmos begins to expand without a big bang, from a region of the size of a Planck volume $V \sim L_*^3$ with the energy density $u \sim u_*$ [18].

Other approaches to anchor Planck's length as a universal elementary length in physical statistics can be tracked in the studies [25] – [30]. Suggestions for the experimental detection of effects due to new physics in the Planck region can be found, e.g., in [31] – [35]. Detailed information on the current state of Lorentz invariance testing is given in [36].

In the following sections, the properties of the new radiation laws (1.3) and (1.4) will be statistically well founded and applied to the general case of ideal quantum gases. The symbols and abbreviations used in this report denote the following quantities:

*General quantities:* $G$ Newton's gravitational constant, $c$ speed of light in vacuum, $h = 2\pi\hbar$ Planck's constant, $k_B$ Boltzmann's constant, $S$ entropy, $E$ energy, $U$ internal energy, $V$ volume, $N$ number of particles, $P$ pressure, $T$ absolute temperature, $\mu$ chemical potential, $p$ momentum, $\omega$ angular frequency, $v = V/N$ specific volume, $g_s$ spin degeneration factor.

*Planck quantities:* Mass $M_* = \sqrt{\hbar c/G}$, length $L_* = GM_*/c^2$, time $t_* = L_*/c$, energy $E_* = M_* c^2$, momentum $P_* = M_* c$, volume $V_* = 2\pi^2 L_*^3$, frequency $\Omega_* = E_*/\hbar$, energy density $u_* = E_*/V_*$, the maximal number of particles in volume $V$ is defined as $N_* = V/V_*$.

## 2 The number of microstates

Consider a thermodynamic system of N non-interacting quantum particles in a fixed volume $V$, which has a total energy $E$ ($N$, $E$ mean values of particle number and energy). Let us therefore consider the motion of a particle picked at random being treated as a quantum-mechanical 1-particle problem and all its energy levels $\varepsilon_k$ being determined. Looking at the overall system of all particles of the ideal quantum gas, we collect a great number $g_k$ of 1-particle states with the energy $\sim \varepsilon_k$ to form a bundle of states

$$|k\rangle \equiv (|k\rangle_1, |k\rangle_2, \cdots\cdots, |k\rangle_{g_k}) \quad , \tag{2.1}$$

which is occupied with a total of $N_k$ particles. In other words, we regard an energy cell $\varepsilon_k$ with $g_k$ states and $N_k$ particles in it. With some justification we can say, then, that this collective state $|k\rangle$ with the energy $\varepsilon_k$ has the „degree of degeneration" $g_k$.

We want to know now how the $N_k$ indistinguishable quantum particles in $|k\rangle$ (at the temperature $T$) are distributed among the $g_k$ states existing there (with the energy $\varepsilon_k$). The fixing of the maximum number of particles allowed to occupy each state, then, yields the various kinds of statistics (Bose, Fermi, or Boltzmann statistics).

Let us use Boltzmann's relation between entropy $S$ and thermodynamic probability $\Omega$ in the form

$$S = k_B \ln \Omega \quad , \tag{2.2}$$

where $\Omega$ is the total number of microstates belonging to our thermodynamic system with the macroscopic parameters „mean energy $E$" and „mean particle number $N$" in the fixed volume $V$. For counting the microstates, let us be informed by the schematic diagram shown in Fig. 1 (see Appendix).

The bundle of states $|k\rangle$ first contains $g_k$ states with the energy $\varepsilon_k$, which are occupied by $N_k$ quantum particles. Let us now distribute the $N_k$ quanta to the $g_k$ states. There will also occur $g_k^{(r)}$ states having an identical occupation number of $r$ quanta. With $r = r_k$ being the maximum occupation number of a state in the $|k\rangle$ bundle, we have $g_k = \sum_{r=0}^{r=r_k} g_k^{(r)}$ as the total number of states and $N_k = \sum_{r=0}^{r=r_k} r \cdot g_k^{(r)}$ as the total number of particles therein. To obtain the number $\Omega_k$ of the microstates at the energy level $\varepsilon_k$, we permute the suitably numbered states $g_k$ together with their respective particle occupations. Because of the indistinguishability of the quantum particles distributed among the states, however, we must take into account that the permutation of states with equal occupation numbers does not create any new microstate, so that, for the number $\Omega_k^0$ of the microstates, there follows

$$\Omega_k^0 = \frac{g_k!}{g_k^{(0)}!\,g_k^{(1)}!\cdots g_k^{(r)}!\cdots g_k^{(r_k)}!} \quad , \quad g_k = \sum_{r=0}^{r=r_k} g_k^{(r)} \; . \tag{2.3}$$

$\Omega_k^0$ stands for the number of microstates as determined in a continuous spacetime manifold by means of common quantum statistics. With this number $\Omega_k^0$ of the microstates, we get the well-known statistics of ideal quantum gases by the usual procedure.

To be able to take into account the hypothesis (1.2) of universally limited quantum leaps, we assume, in addition, that each state occurs in the bundle of states $|k\rangle$ with a probability $q_k$ that is equal for all $g_k$ states in $|k\rangle$. A statement equivalent to this is that the energy cell $\varepsilon_k$ with all its $g_k$ states now no longer occurs with certainty, but only with the probability $q_k(\varepsilon_k)$. For illustration, this can be interpreted in such a way that each of the $g_k$ states shows an added $\lambda_k$-fold degeneration ($\lambda_k = 1/q_k$), which is impressed on the thermodynamic system by a fluctuating microstructure of the spacetime manifold.

The total number of possible states in the $|k\rangle$-bundle is thus increased to $G_k = g_k \cdot \lambda_k$ (Fig. 1). The number $\Omega_k$ of the microstates now becomes

$$\Omega_k = \frac{G_k!}{g_k^{(0)}!\,g_k^{(1)}!\,g_k^{(2)}!\cdots g_k^{(r)}!\cdots g_k^{(r_k)}!} \cdot q_k^{N_k} \quad , \quad G_k = \sum_{r=0}^{r_k} g_k^{(r)} \quad , \quad N_k = \sum_{r=0}^{r_k} r \cdot g_k^{(r)} \; , \tag{2.4}$$

with the factor $q_k^{N_k}$ describing the probability of finding just $N_k$ particles in the bundle of states $|k\rangle$ with the energy $\varepsilon_k$. The fundamental quantum-statistical assumption of the a priori equiprobability of all microstates of the overall system is thus restricted: Only the $g_k$ microstates within the bundle of states $|k\rangle$ have the same initial probability $q_k(\varepsilon_k)$. Via this parameter $q_k$ we will, below, be able to take into account the hypothesis (1.2) of universally limited quantum leaps in the equations of state of the quantum gases. If $q_k = 1/\lambda_k = 1$, the fluctuations of the numbers of state $g_k$ vanish, and we have $\Omega_k = \Omega_k^0$ again.

With (2.4), therefore, the total number of microstates of our model of ideal quantum gases is

$$\Omega = \prod_k \Omega_k \quad , \quad \Omega_k = \frac{G_k!}{\prod_{r=0}^{r_k} g_k^{(r)}!} \cdot q_k^{N_k} \; . \tag{2.5}$$

## 3 Entropy maximization

We find the maximum of the entropy $S$ under certain constraints by means of Lagrange's multiplier method. With (2.5), we get

$$S = k_B \ln \Omega = k_B \ln \prod_k \Omega_k = k_B \sum_k \ln \Omega_k = \sum_k S_k \ . \tag{3.1}$$

We select the following constraints of entropy maximization:

$$N = \sum_k N_k = \sum_k \sum_{r=0}^{r_k} r\, g_k^{(r)} \ , \tag{3.2}$$

$$E = \sum_k E_k = \sum_k N_k \varepsilon_k = \sum_k \sum_{r=0}^{r_k} r\, g_k^{(r)} \varepsilon_k \ , \tag{3.3}$$

$$G_k = \sum_{r=0}^{r_k} g_k^{(r)} \ . \tag{3.4}$$

According to Schmutzer [19], one can arbitrarily choose the sufficiently large total number $G_k$ of the states in the $\varepsilon_k$ range, which we have already made by specification of the structure $G_k = g_k \cdot \lambda_k$.

Now we vary the number of states $g_k^{(r)}$ occupied by $r$ particles. With Lagrange's multipliers $k_B \cdot \alpha$, $k_B \cdot \beta$, $k_B \cdot \gamma_k$, the complete variation problem reads

$$\delta \sum_k \left\{ \ln \Omega_k - \alpha \sum_{r=0}^{r_k} r g_k^{(r)} - \beta \sum_{r=0}^{r_k} r \varepsilon_k g_k^{(r)} - \gamma_k \sum_{r=0}^{r_k} g_k^{(r)} \right\} = 0 \ , \tag{3.5}$$

$$\sum_k \sum_{r=0}^{r_k} \left\{ \frac{\partial \ln \Omega_k}{\partial g_k^{(r)}} - (\alpha + \beta \varepsilon_k) \cdot r - \gamma_k \right\} \delta g_k^{(r)} = 0 \ . \tag{3.6}$$

With (2.5) and Stirling's approximation $\ln(x!) = x(\ln x - 1)$, we obtain

$$\ln \Omega_k = \ln(G_k!) - \sum_{r=0}^{r_k} \ln(g_k^{(r)}!) + N_k \ln q_k \ , \tag{3.7}$$

$$\ln \Omega_k = G_k (\ln G_k - 1) - \sum_{r=0}^{r_k} g_k^{(r)} (\ln g_k^{(r)} - 1) + (\ln q_k) \cdot \sum_{r=0}^{r_k} r g_k^{(r)} \ , \tag{3.8}$$

with $N_k = \sum_{r=0}^{r=r_k} r \cdot g_k^{(r)}$ being used in the last term. Thus we get

$$\frac{\partial \ln \Omega_k}{\partial g_k^{(r)}} = \sum_{r=0}^{r_k} \left( 0 - \ln g_k^{(r)} + r \ln q_k \right) \ . \tag{3.9}$$

With (3.9) substituted in (3.6) we get the result of the variation

$$\sum_k \sum_{r=0}^{r_k} \left\{ -\ln g_k^{(r)} + r \cdot \ln q_k - (\alpha + \beta \varepsilon_k) \cdot r - \gamma_k \right\} \delta g_k^{(r)} = 0 \ . \tag{3.10}$$

Since all variations $\delta g_k^{(r)}$ may now be performed independently of each other, the content of the braces must vanish for each individual $r$. This leads to

$$g_k^{(r)} = \frac{1}{\exp(\gamma_k)} \left\{ q_k \exp(-(\alpha + \beta \varepsilon)) \right\}^r \ . \tag{3.11}$$

After determining the parameter $\gamma_k$ from the constraint (3.4), we have the provisional result

$$g_k^{(r)} = G_k \frac{\left\{ q_k \cdot \exp(-(\alpha + \beta \varepsilon_k)) \right\}^r}{\sum_{r=0}^{r_k} \left\{ q_k \cdot \exp(-(\alpha + \beta \varepsilon_k)) \right\}^r} \quad , \quad w_k^{(r)} = \frac{g_k^{(r)}}{G_k} = q_k \cdot \frac{g_k^{(r)}}{g_k} \ , \tag{3.12}$$

since $G_k = \lambda_k \cdot g_k = g_k / q_k$. Thus, $w_k^{(r)}$ is the probability of finding a state occupied by r quantum particles in the energy level $\varepsilon_k$.

## 4 Transition to thermodynamics

To identify the parameters $\alpha$ and $\beta$ in (3.11), we include (3.12) and compute the entropy of the bundle of states $|k\rangle$,

$$S_k = k_B \ln \Omega_k \ . \tag{4.1}$$

With (3.8) and (3.12), after lengthy but simple conversions, there follows

$$S_k = k_B \left\{ \ln Z_k + (\alpha + \beta \varepsilon_k) N_k \right\} \ , \tag{4.2}$$

where $Z_k$ is the grand canonical partition function for the $|k\rangle$-bundle,

$$Z_k = \left\{ \sum_{r=0}^{r_k} \left[ q_k \exp(-(\alpha + \beta \varepsilon)) \right]^r \right\}^{G_k} \ . \tag{4.3}$$

For the total system of the non-interacting quantum gas, the grand canonical potential $\Phi$ and the partition function $Z$ are defined by

$$\Phi = -k_B T \ln Z \quad , \quad Z = \prod_k Z_k \quad , \quad \Phi = \sum_k \Phi_k \ . \tag{4.4}$$

Entropy $S$, mean energy $E$ and mean particle number $N$ are simply

$$S = \sum_k S_k \quad , \quad E = \sum_k E_k \quad , \quad N = \sum_k N_k \quad . \tag{4.5}$$

Then, with summation over $k$ from (4.2) and multiplication with the temperature $T$, we get

$$TS = k_B T \{\ln Z + \alpha N + \beta E\} \quad . \tag{4.6}$$

By comparison with the Gibbs-Duhem relation of equilibrium thermodynamics

$$TS = +PV - \mu N + E \tag{4.7}$$

the grand canonical potential $\Phi$ as well as $\alpha$ and $\beta$ are defined,

$$\Phi = -PV = -k_B T \ln Z \quad , \quad \alpha = -\frac{\mu}{k_B T} \quad , \quad \beta = \frac{1}{k_B T} \quad . \tag{4.8}$$

With $\Phi = \Phi(T,V,\mu)$ as potential and $\eta = \exp(-\alpha) = \exp(\beta\mu)$ as fugacity, there follow, as usual, the three major equations of state (EoS),

$$PV = k_B T \ln Z \qquad \text{thermal EoS} \quad , \tag{4.9}$$

$$E = -\frac{\partial}{\partial \beta}(\ln Z)\bigg|_{\eta,V} \qquad \text{caloric EoS} \quad , \tag{4.10}$$

$$N = \eta \frac{\partial}{\partial \eta}(\ln Z)\bigg|_{\beta,V} \qquad \text{chemical EoS} \quad . \tag{4.11}$$

The fluctuations of energy and particle number about their mean values $E$ and $N$ are known to be computable from

$$\overline{(\Delta E)^2} = -\frac{\partial E}{\partial \beta} = +\frac{\partial^2 (\ln Z)}{\partial \beta^2}\bigg|_{\eta,V} \quad , \tag{4.12}$$

$$\overline{(\Delta N)^2} = +\eta \frac{\partial N}{\partial \eta} = +\left(\eta \frac{\partial}{\partial \eta}\right)\left(\eta \frac{\partial}{\partial \eta}\right)(\ln Z)\bigg|_{\beta,V} \quad . \tag{4.13}$$

If the 1-particle energies $\varepsilon_k$ are sufficiently close to each other (which is certainly the case for macroscopic volumes), we can proceed to the integral form of the functions of state. Let this be exemplified by the grand canonical potential $\Phi$, from which all other functions of state can be derived (cf. 4.7 to 4.11). According to (4.3) and (4.4),

$$\Phi = -k_B T \ln Z = -k_B T \sum_k \ln Z_k \;, \tag{4.14}$$

$$\Phi = -k_B T \sum_k G_k \ln \left\{ \sum_{r=0}^{r_k} \left[ q_k \exp(-(\alpha + \beta \varepsilon_k)) \right]^r \right\} . \tag{4.15}$$

Considering $G_k = g_k \cdot \lambda_k = g_k / q_k$, we finally have

$$\Phi = -k_B T \sum_k \left( \frac{g_k}{q_k} \right) \ln \left\{ \sum_{r=0}^{r_k} \left[ q_k \exp(-(\alpha + \beta \varepsilon_k)) \right]^r \right\} . \tag{4.16}$$

Transition to the integral is effected by

$$\sum_k \;\Rightarrow\; \frac{1}{\Delta^3 k} \int d^3 k \;, \qquad \Delta^3 k = \left( \frac{2\pi}{L} \right)^3 = \frac{(2\pi)^3}{V} \;, \tag{4.17}$$

wherein we choose a great reference length $L$ compared to the wavelengths of the quantum particles. With $d^3 k = 4\pi k^2 dk$, then, the complete transition rule reads

$$\sum_k A_k \;\Rightarrow\; \frac{4\pi V}{(2\pi)^3} \int dk \cdot k^2 A(k) \;. \tag{4.18}$$

Applied to (4.16) and with, e.g., $g_k = g_s$ (spin degrees of freedom of the quantum particles),

$$\Phi = -g_s (k_B T) \frac{4\pi V}{(2\pi)^3} \int_0^{k_*} \frac{k^2}{q(k)} \ln \left\{ \sum_{r=0}^{r_k} \{ q(k) \cdot \exp[-(\alpha + \beta \varepsilon(k))] \}^r \right\} dk \;. \tag{4.19}$$

With $\vec{p} = \hbar \vec{k}$, converted in a momentum integral, the grand canonical potential $\Phi$ then adopts the form of

$$\Phi = -g_s (k_B T) \frac{4\pi V}{h^3} \int_0^{P_*} \frac{p^2}{q(p)} \ln \left\{ \sum_{r=0}^{r_k} \{ q(p) \cdot \exp[-(\alpha + \beta \varepsilon(p))] \}^r \right\} dp \;. \tag{4.20}$$

Because of $\Phi = -k_B T \ln Z$, one can read from (4.20) that

$$\ln Z = g_s \frac{4\pi V}{h^3} \int_0^{P_*} \frac{p^2}{q(p)} \ln \left\{ \sum_{r=0}^{r_k} \{ q(p) \cdot \exp[-(\alpha + \beta \varepsilon(p))] \}^r \right\} dp \;. \tag{4.21}$$

The upper integration limit $P_* = \hbar k_* = E_* / c$ corresponds to the existence of a maximum momentum transfer between the quantum particles following from the hypothesis (1.2).

These form of the grand canonical potential (4.20) reveals the effect of introduction of the $q_k$ probability factor in the ansatz (2.4) for the total number $\Omega_k$ of the microstates: The factor $q_k$ in $q_k^{N_k}$ generates the $q(p)$ factor preceding the exponential function in $\Phi$, and the factor $1/q_k$ in $G_k = g_k \cdot \lambda_k = g_k/q_k$ gives rise to the factor $1/q(p)$ preceding the logarithm in the potential $\Phi$. The latter factor, in particular, in the measure of integration

$$\frac{4\pi p^2 dp\, dV}{h^3 q(p)} = \frac{d^3 p\, d^3 q}{h^3 q(p)} \quad , \quad q(p) = \left(1 - \frac{\varepsilon(p)}{E_*}\right), \tag{4.22}$$

points (besides $h$) to another now energy-dependent partition of the phase-space volume (see also [28]). The form of $q(p)$ is determined in the following chapter 5.

## 5 Fixing the probability parameter $q_k$

Because of the universal importance of the quantum leap limit according to hypothesis (1.2), we can set the value of the parameter $q_k$ for a specific quantum gas, viz. photon gas. It is for this case that we already derived the law of distribution of energy by another, direct and elementary method [17]. With the aid of Einstein's laser model and taking into account the hypothesis (1.2) for the photon gas, we obtained the following distribution formula for the mean (spectral) energy $\bar{\varepsilon}_k(T)$ per state in the bundle of states $|k\rangle$:

$$\bar{\varepsilon}_k(T) = \frac{\varepsilon_k}{\exp\left(\dfrac{\varepsilon_k}{k_B T}\right) - \left(1 - \dfrac{\varepsilon_k}{E_*}\right)} \quad . \tag{5.1}$$

Here, $|k\rangle$ denotes all states with the wavenumber $k = \omega/c$ and the energy $\varepsilon_k = \hbar k \cdot c = \hbar\omega$.

We now compute the mean energy $\bar{\varepsilon}_k(T)$ by the statistical method described in chapter 4. According to the formulas (4.3), (4.4), (4.5) and (4.10),

$$E = \sum_k E_k = -\frac{\partial}{\partial \beta}(\ln Z)\bigg|_{\eta,V} = -\sum_k \frac{\partial}{\partial \beta}(\ln Z_k)\bigg|_{\eta,V}, \tag{5.2}$$

$$E_k = -\frac{\partial}{\partial \beta}(\ln Z_k)\bigg|_{\eta,V} \quad , \quad Z_k = \left\{\sum_{r=0}^{r_k}\left[q_k \exp(-(\alpha + \beta \varepsilon_k))\right]^r\right\}^{G_k}. \tag{5.3}$$

If the ideal quantum gas is a photon gas with $\mu = \alpha = 0$, each state in $|k\rangle$ may be occupied by an infinite number of quanta, i.e., the maximum occupation number is $r_k \to \infty$. Then, with $q_k < 1$, $G_k = g_k \cdot \lambda_k = g_k/q_k$ and $\beta = 1/k_B T$, there results

$$Z_k = \left\{ \sum_{r=0}^{\infty} \left[ q_k \exp(-\beta \varepsilon_k) \right]^r \right\}^{G_k} = \left\{ 1 - q_k \exp(-\beta \varepsilon_k) \right\}^{-\frac{g_k}{q_k}} \quad , \tag{5.4}$$

$$E_k = \frac{\partial}{\partial \beta} \ln Z_k = g_k \cdot \frac{\varepsilon_k}{\exp(\beta \varepsilon_k) - q_k} \quad . \tag{5.5}$$

$E_k$ is the mean energy of the bundle of states $|k\rangle$ at the temperature $T$, equally distributed among the $g_k$ states. Our interest is focused on the mean energy $\bar{\varepsilon}_k$ referred to one of the $g_k$ states,

$$\bar{\varepsilon}_k = \frac{\varepsilon_k}{\exp\left(\frac{\varepsilon_k}{k_B T}\right) - q_k} \quad . \tag{5.6}$$

By comparison with the distribution formula (5.1) derived from the laser model mentioned above we deduce the value of the probability parameter $q_k$, according to hypothesis (1.2) now valid for all distribution formulas of ideal quantum gases,

$$q_k = 1 - \frac{\varepsilon_k}{E_*} \quad . \tag{5.7}$$

Let the physical significance of this probability parameter $q_k$ get clear to us again: Each of the $g_k$ states occurring in the bundle of states $|k\rangle$ with the energy $\varepsilon_k$ exists there only with a probability of $q_k(\varepsilon_k) = 1 - \varepsilon_k / E_*$. Such an assumption goes beyond the conventional quantum-statistical fundamentals, in which the states exist exactly and invariably (i.e. with the probability $q_k = 1$). Here, this case occurs only with $E_* \to \infty$, i.e., in quantum statistics based on a quantum theory in which the height of the energetic quantum leap is unlimited. Our presumption is that, in a perfect theory of quantum gravitation with a fluctuating spacetime structure, the model of an unsharp, fluctuating number of states might occur.

## 6 Special ideal quantum gases

It is known that the specification of a maximum number $r_k$ of the particles located in each of the states of the bundle of states $|k\rangle$ results in different statistics. Fermi statistics appears with a maximum occupation number $r_k = 1$, Bose statistics with $r_k \to \infty$. In the following chapter, we do this concretely and after that regard some limit cases. Our starting point is the grand canonical potential $\Phi$ of an ideal quantum gas, which, according to (4.3) – (4.8) has the form

$$\Phi = -k_B T \ln Z = -PV \quad , \quad Z = \prod_k Z_k \quad , \tag{6.1}$$

$$Z_k = \left\{\sum_{r=0}^{r_k}\left[q_k \exp(-(\alpha + \beta\varepsilon_k))\right]^r\right\}^{G_k} \quad , \quad G_k = \frac{g_k}{q_k} \; . \tag{6.2}$$

In the integral form we received in (4.20) for the grand canonical potential the momentum integral ($r_k = 1$: fermions, $r_k \to \infty$: bosons)

$$\Phi = -g_s \frac{4\pi V}{h^3} k_B T \int_0^{P_*} \frac{p^2}{q(p)} \ln\left\{\sum_{r=0}^{r_k}\{q(p)\exp[-(\alpha + \beta\varepsilon(p))]\}^r\right\} dp \tag{6.3}$$

and hence, from (6.1), for the logarithm of the grand canonical partitions function

$$\ln Z = g_s \frac{4\pi V}{h^3} \int_0^{P_*} \frac{p^2}{q(p)} \ln\left\{\sum_{r=0}^{r_k}\{q(p)\exp[-(\alpha + \beta\varepsilon(p))]\}^r\right\} dp \; . \tag{6.4}$$

By means of (4.9) – (4.11) and considering (5.7) in the continuous form

$$q(p) = 1 - \frac{\varepsilon(p)}{E_*} \tag{6.5}$$

there follow all new equations of state of ideal quantum gases.

**6.1 Fermi statistics**

Quantum systems in which each state may be occupied by maximally one particle are subject to Fermi statistics. Let us now evaluate the statistical principles of the previous chapters with the maximum occupation number $r_k = 1$. In this case, the partition function $Z_k$ of the bundle of states $|k\rangle$ appearing in (6.2) takes the form

$$Z_k = \left\{\sum_{r=0}^{r=r_k=1}\left[q_k \exp(-(\alpha + \beta\varepsilon_k))\right]^r\right\}^{G_k} = \left\{1 + q_k\, \eta\, \exp(-\beta\varepsilon_k)\right\}^{\left(\frac{g_k}{q_k}\right)} \; . \tag{6.6}$$

For the mean energy $E_k$ and mean particle number $N_k$ in the bundle of states $|k\rangle$, we compute

$$E_k = -\frac{\partial}{\partial \beta}(\ln Z_k)\bigg|_\eta = g_k \cdot \frac{\varepsilon_k}{\exp(\alpha + \beta\varepsilon_k) + q_k} \quad , \tag{6.7}$$

$$N_k = \eta \frac{\partial}{\partial \eta}(\ln Z_k)\bigg|_\beta = g_k \cdot \frac{1}{\exp(\alpha + \beta\varepsilon_k) + q_k} \; . \tag{6.8}$$

For the mean energy $\bar{\varepsilon}_k = E_k/g_k$ and the mean particle number $\bar{n}_k = N_k/g_k$, with reference to one of the $g_k$ states, we can then, with $q_k$ from (5.7) and $\alpha, \beta$ from (4.8), obtain

$$\bar{\varepsilon}_k = \frac{\varepsilon_k}{\exp\left(\frac{\varepsilon_k - \mu}{k_B T}\right) + \left(1 - \frac{\varepsilon_k}{E_*}\right)} \quad , \quad \bar{n}_k = \frac{1}{\exp\left(\frac{\varepsilon_k - \mu}{k_B T}\right) + \left(1 - \frac{\varepsilon_k}{E_*}\right)} \tag{6.9}$$

with

$$\bar{\varepsilon}_k = \bar{n}_k \cdot \varepsilon_k \, . \tag{6.10}$$

With (6.4) for the Fermi statistics and the derivatives (4.9) – (4.11) we immediately obtain the integral form of the thermal, caloric and chemical equations of state as momentum integrals

$$P = g_s \frac{4\pi}{h^3}(k_B T) \int_0^{P_*} \frac{p^2}{\left(1 - \frac{\varepsilon(p)}{E_*}\right)} \ln\left\{1 + \left(1 - \frac{\varepsilon(p)}{E_*}\right)\exp\left[-\left(\frac{\varepsilon(p) - \mu}{k_B T}\right)\right]\right\} dp \, , \tag{6.11}$$

$$E = g_s \frac{4\pi V}{h^3} \int_0^{P_*} \frac{\varepsilon(p)}{\exp\left(\frac{\varepsilon(p) - \mu}{k_B T}\right) + \left(1 - \frac{\varepsilon(p)}{E_*}\right)} p^2 dp \, , \tag{6.12}$$

$$N = g_s \frac{4\pi V}{h^3} \int_0^{P_*} \frac{1}{\exp\left(\frac{\varepsilon(p) - \mu}{k_B T}\right) + \left(1 - \frac{\varepsilon(p)}{E_*}\right)} p^2 dp \, , \tag{6.13}$$

with integration extending up to the Planck momentum $P_* = E_*/c$.

## 6.2 Bose statistics

Quantum systems in which each state may be occupied by any number of particles are subject to Bose statistics. To implement this, we now evaluate the statistical principles of the previous chapters with the maximum occupation number $r_k \to \infty$. In this case, the partition function $Z_k$ of the bundle of states $|k\rangle$ appearing in (6.1) takes the form

$$Z_k = \left\{ \sum_{r=0}^{r=r_k \to \infty} \left[q_k \exp(-(\alpha + \beta\varepsilon_k))\right]^r \right\}^{G_k} = \left\{\frac{1}{1 - q_k \eta \exp(-\beta\varepsilon_k)}\right\}^{\left(\frac{g_k}{q_k}\right)} , \tag{6.14}$$

with the fugacity $\eta = \exp(-\alpha)$.

However, because of the convergence of the row sum, it is always necessary that

$$q_k \eta < 1 \tag{6.15}$$

which can be satisfied, e.g., by the two single requirements

$$\alpha + \beta \varepsilon_k > 1 \quad , \quad q_k < 1 \tag{6.16}$$

With $\alpha = -\mu/k_B T$, $\beta = 1/k_B T$, $q_k = 1 - \varepsilon_k/E_*$, we can combine the inequalities (6.16) into the double form

$$\mu < \varepsilon_k < E_* \tag{6.17}$$

As the smallest energy value for bosons can be $\varepsilon_k = 0$, there applies $\mu < 0$ for them. For fermions, the maximum $\mu$-value is exactly the Fermi energy $\mu(T=0) = \varepsilon_F$. The maximum value of this, in turn, is Planck's energy $\varepsilon_F = E_*$, because of the hypothesis (1.2). Therefore, the inequality (6.17) should apply to all ideal quantum gases.

For the mean energy $E_k$ and the mean particle number $N_k$ in the $|k\rangle$ range we get, with the partition function (6.14),

$$E_k = -\frac{\partial}{\partial \beta}(\ln Z_k)\bigg|_\eta = g_k \cdot \frac{\varepsilon_k}{\exp(\alpha + \beta \varepsilon_k) - q_k} \quad , \tag{6.18}$$

$$N_k = \eta \frac{\partial}{\partial \eta}(\ln Z_k)\bigg|_\beta = g_k \cdot \frac{1}{\exp(\alpha + \beta \varepsilon_k) - q_k} \quad , \tag{6.19}$$

Here, the mean energies $\bar{\varepsilon}_k = E_k/g_k$ and mean particle numbers $\bar{n}_k = N_k/g_k$, referred to one of the $g_k$ states, with $q_k$ from (5.7) and $\alpha, \beta$ from (4.8), have the form

$$\bar{\varepsilon}_k = \frac{\varepsilon_k}{\exp\left(\frac{\varepsilon_k - \mu}{k_B T}\right) - \left(1 - \frac{\varepsilon_k}{E_*}\right)} \quad , \quad \bar{n}_k = \frac{1}{\exp\left(\frac{\varepsilon_k - \mu}{k_B T}\right) - \left(1 - \frac{\varepsilon_k}{E_*}\right)} \tag{6.20}$$

Here again,

$$\bar{\varepsilon}_k = \bar{n}_k \cdot \varepsilon_k . \tag{6.21}$$

With (6.4) applied to Bose statistics and with the derivatives (4.9) – (4.11), pressure $P$, energy $E$ and particle number $N$ can be given in an integral form:

$$P = -g_s \frac{4\pi}{h^3}(k_B T) \int_0^{P_*} \frac{p^2}{\left(1 - \frac{\varepsilon(p)}{E_*}\right)} \ln\left\{1 - \left(1 - \frac{\varepsilon(p)}{E_*}\right)\exp\left[-\left(\frac{\varepsilon(p) - \mu}{k_B T}\right)\right]\right\} dp \quad , \tag{6.22}$$

$$E = g_s \frac{4\pi V}{h^3} \int_0^{P_*} \frac{\varepsilon(p)}{\exp\left(\frac{\varepsilon(p)-\mu}{k_B T}\right) - \left(1 - \frac{\varepsilon(p)}{E_*}\right)} p^2 dp \quad , \tag{6.23}$$

$$N = g_s \frac{4\pi V}{h^3} \int_0^{P_*} \frac{1}{\exp\left(\frac{\varepsilon(p)-\mu}{k_B T}\right) - \left(1 - \frac{\varepsilon(p)}{E_*}\right)} p^2 dp \quad . \tag{6.24}$$

Remark: If we regard the inequality (6.17) as universally true, there obviously exist in nature three fundamental physical quantities with a universally limited range: signal propagation ($0 < v < c$), action ($\hbar < W < \infty$), and chemical potential ($-\infty < \mu < E_*$).

## 6.3 Fluctuations

From the fluctuations of thermodynamic quantities about their mean values one can draw interesting conclusions. Let us compute the mean square fluctuation of the energy $\overline{(\Delta E_k)^2}$ in the range of states $|k\rangle$ by formula (4.12) for a Fermi and a Bose gas. We proceed from the formulas (6.7) or (6.18), respectively, for the mean energy $E_k$ of the respective gas.

B o s o n s :

From (6.18) we use

$$E_k = g_k \cdot \frac{\varepsilon_k}{\exp(\alpha + \beta \varepsilon_k) - q_k} = g_k \overline{\varepsilon}_k \quad , \tag{6.25}$$

$$\overline{(\Delta E_k)^2} = -\frac{\partial E_k}{\partial \beta} = -\frac{\partial}{\partial \beta}\left\{\frac{g_k \varepsilon_k}{\exp(\alpha + \beta \varepsilon_k) - q_k}\right\} = \frac{1}{g_k} E_k^2 \cdot \exp(\alpha + \beta \varepsilon_k) \quad . \tag{6.26}$$

If the exponential function $\exp(\alpha + \beta \varepsilon_k)$ is, by means of (6.25), again expressed by the mean energy $E_k$ and if we consider $E_k = g_k \overline{\varepsilon}_k$, there follows, for the relative mean square of fluctuation,

$$\frac{\overline{(\Delta E_k)^2}}{E_k^2} = \frac{1}{g_k}\left[\frac{g_k \varepsilon_k}{E_k} + q_k\right] = \frac{1}{E_k}[\varepsilon_k + q_k \overline{\varepsilon}_k] \quad . \tag{6.27}$$

If we also note the additivity of the square fluctuations for extensive quantities, then the fluctuation of the energy in the $|k\rangle$ range is $\overline{(\Delta E_k)^2} = g_k \overline{(\Delta \varepsilon_k)^2}$. Referred to one of the $g_k$ states the fluctuation formula for the Bose gas is now

$$\frac{\overline{(\Delta \varepsilon_k)^2}}{\overline{\varepsilon}_k^2} = \frac{1}{\overline{\varepsilon}_k}\{\varepsilon_k + q_k \overline{\varepsilon}_k\} = \frac{1}{\overline{\varepsilon}_k}\left\{\varepsilon_k - \frac{\varepsilon_k \overline{\varepsilon}_k}{E_*} + \overline{\varepsilon}_k\right\} \ . \tag{6.28}$$

This result is obviously good interpretable. The two positive terms in (6.28) correspond to the known fluctuation shares in the ideal Bose gas ($E_* \to \infty$). The first term originates from the particle character of the quantum gas: If we substitute $\overline{\varepsilon}_k = \overline{n}_k \varepsilon_k$, this expression is proportional to $1/\overline{n}_k$, as known from the classical ideal particle gas. In the third term, the absolute fluctuation $\sqrt{\overline{(\Delta \varepsilon_k)^2}}$ of the energy is as big as the mean energy $\overline{\varepsilon}_k$ itself, as one would expect it from a „wave gas". The new, middle term can, however, satisfy two functions:

If we keep the 1-particle-energy $\varepsilon_k$ fixed and increase the temperature $T$ of the Bose gas and, thus, the mean energy $\overline{\varepsilon}_k$, we reduce the particle character of the gas,

$$\frac{\overline{(\Delta \varepsilon_k)^2}}{\overline{\varepsilon}_k^2} = \frac{1}{\overline{\varepsilon}_k}\left\{\varepsilon_k\left(1 - \frac{\overline{\varepsilon}_k}{E_*}\right) + \overline{\varepsilon}_k\right\} \xrightarrow{T \to \infty} 1 \ . \tag{6.29}$$

In passing to the limit $\lim_{T \to \infty} \overline{\varepsilon}_k = E_*$ (see formula (6.20)), the particle share of the fluctuation even vanishes, leaving a pure wave gas.

If, on the other hand, we keep the temperature $T$ of the Bose gas fixed and increase the particle energy $\varepsilon_k$ of the single particle, we reduce the wave character of the gas,

$$\frac{\overline{(\Delta \varepsilon_k)^2}}{\overline{\varepsilon}_k^2} = \frac{1}{\overline{\varepsilon}_k}\left\{\varepsilon_k + \left(1 - \frac{\varepsilon_k}{E_*}\right)\overline{\varepsilon}_k\right\} \xrightarrow{\varepsilon_k \to E_*} \frac{\varepsilon_k}{\overline{\varepsilon}_k} = \frac{1}{\overline{n}_k} \ . \tag{6.30}$$

In passing to the limit $\varepsilon_k \to E_*$, the wave share of the fluctuation vanishes, leaving a pure particle gas.

F e r m i o n s :

In the case of fermions, the analogous procedure lead to similar formulas for the relative fluctuation of energy in the bundle of states $|k\rangle$. From (6.7),

$$E_k = g_k \cdot \frac{\varepsilon_k}{\exp(\alpha + \beta \varepsilon_k) + q_k} = g_k \overline{\varepsilon}_k \ , \tag{6.31}$$

$$\overline{(\Delta E_k)^2} = -\frac{\partial E_k}{\partial \beta} = -\frac{\partial}{\partial \beta}\left\{\frac{g_k \varepsilon_k}{\exp(\alpha + \beta \varepsilon_k) + q_k}\right\} = \frac{1}{g_k} E_k^2 \cdot \exp(\alpha + \beta \varepsilon_k) \ . \tag{6.32}$$

By means of (6.31), the exponential function $\exp(\alpha + \beta \varepsilon_k)$ is expressed by the mean energy $E_k$. Taking into account $E_k = g_k \overline{\varepsilon}_k$, we get

$$\frac{\overline{(\Delta E_k)^2}}{E_k^2} = \frac{1}{g_k}\left[\frac{g_k \varepsilon_k}{E_k} - q_k\right] = \frac{1}{E_k}[\varepsilon_k - q_k \overline{\varepsilon}_k] \quad . \tag{6.33}$$

With $\overline{(\Delta E_k)^2} = g_k \overline{(\Delta \varepsilon_k)^2}$ and referred to one of the $g_k$ states, the fluctuation formula for the Fermi gas is

$$\frac{\overline{(\Delta \varepsilon_k)^2}}{\overline{\varepsilon}_k^2} = \frac{1}{\overline{\varepsilon}_k}\{\varepsilon_k - q_k \overline{\varepsilon}_k\} = \frac{1}{\overline{\varepsilon}_k}\left\{\varepsilon_k + \frac{\varepsilon_k \overline{\varepsilon}_k}{E_*} - \overline{\varepsilon}_k\right\} \quad . \tag{6.34}$$

Here again, the first and third terms (6.34) originate from the known fluctuation formula of the ideal Fermi gas. With the new, middle term, we again discuss the interplay between the wave and the particle character of the Fermi gas:

If we keep the temperature $T$ of the Fermi gas constant while increasing the particle energy $\varepsilon_k$ of the single particle, we reduce the wave character of the gas,

$$\frac{\overline{(\Delta \varepsilon_k)^2}}{\overline{\varepsilon}_k^2} = \frac{1}{\overline{\varepsilon}_k}\left[\varepsilon_k - \left(1 - \frac{\varepsilon_k}{E_*}\right)\overline{\varepsilon}_k\right] \xrightarrow{\varepsilon_k \to E_*} \frac{\varepsilon_k}{\overline{\varepsilon}_k} = \frac{1}{n_k} \quad . \tag{6.35}$$

In passing to the limit $\varepsilon_k \to E_*$, the wave share of the fluctuation vanishes, leaving a pure particle gas.

On the other hand, if we keep the 1-particle-energy constant while increasing the temperature $T$ of the Fermi gas and, thus, the mean energy $\overline{\varepsilon}_k$, we reduce the particle character of the gas,

$$\frac{\overline{(\Delta \varepsilon_k)^2}}{\overline{\varepsilon}_k^2} = \frac{1}{\overline{\varepsilon}_k}\left[\left(1 + \frac{\overline{\varepsilon}_k}{E_*}\right)\varepsilon_k - \overline{\varepsilon}_k\right] = \left[\frac{\varepsilon_k}{\overline{\varepsilon}_k} + \frac{\varepsilon_k}{E_*} - 1\right] \xrightarrow{T \to \infty} 1 \quad . \tag{6.36}$$

because in passing to the limit $T \to \infty$ (see formula (6.9))

$$\lim_{T \to \infty} \overline{\varepsilon}_k = \frac{\varepsilon_k}{2 - \frac{\varepsilon_k}{E_*}} \quad , \quad \lim_{T \to \infty}\left(\frac{\varepsilon_k}{\overline{\varepsilon}_k}\right) = 2 - \frac{\varepsilon_k}{E_*} \quad . \tag{6.37}$$

The particle share of the fluctuation (6.36) vanishes altogether, leaving a pure wave gas.

Generally, it can be seen that there is a change in the behaviour of the fluctuations for ideal quantum gases, which has its cause in the appearance of the $q_k$-term in the distribution formulas (6.9) or (6.20), respectively. If, e.g., the 1-particle energy is increased up to $\varepsilon_k \to E_*$ while the temperature $T$ is kept constant, $q_k = (1 - \varepsilon_k/E_*) \to 0$ vanishes, and both distribution formulas turn into the common Boltzmann distribution for a classical ideal particle gas. Then follows of course the corresponding fluctuation formula proportional to the known term $(1/\overline{n}_k)$. It should be noted further that here we are making statements about statistical properties of a quantum system: Depending the selection of the parameters $T$ or $\varepsilon_k$ of the quantum system, we can influence its wave or particle aspect, respectively. In limit cases, we even can suppress one of the two. This is not possible in conventional quantum statistics! One can assume that a complete theory of quantum gravitation, in which Planck quantities such as $E_*$ or $L_*$ should play a fundamental part, will enlighten us about this (cf. also [23]).

## 7 Properties of ideal quantum gases

We want to specify the most important properties of non-relativistic and relativistic ideal quantum gases. Let us investigate the equations of state and the behaviour at low and high temperatures if the term $q = 1 - (\varepsilon/E_*)$ occurs in the distribution functions.

We start from the thermal and caloric equations of state for bosonic and fermionic ideal quantum gases as formulated in formulas (6.11) – (6.13) and (6.22) –(6.24). Condensed, they read

$$P = \pm g_s \frac{4\pi}{h^3} k_B T \int_0^{P_*} \frac{p^2}{\left(1 - \frac{\varepsilon(p)}{E_*}\right)} \ln\left\{1 \pm \left(1 - \frac{\varepsilon(p)}{E_*}\right) \exp\left[-\left(\frac{\varepsilon(p) - \mu}{k_B T}\right)\right]\right\} dp , \qquad (7.1)$$

$$E = g_s \frac{4\pi}{h^3} V \int_0^{P_*} \frac{\varepsilon(p)}{\exp\left(\frac{\varepsilon(p) - \mu}{k_B T}\right) \pm \left(1 - \frac{\varepsilon(p)}{E_*}\right)} p^2 dp , \qquad (7.2)$$

$$N = g_s \frac{4\pi}{h^3} V \int_0^{P_*} \frac{1}{\exp\left(\frac{\varepsilon(p) - \mu}{k_B T}\right) \pm \left(1 - \frac{\varepsilon(p)}{E_*}\right)} p^2 dp , \qquad (7.3)$$

with the upper sign applying to fermions and the lower one to bosons.

## 7.1 Non-relativistic ideal quantum gases

We start in (7.1) – (7.3) with the exact relativistic energy-momentum relation. For sufficiently small momenta $p \ll m_0 c$ (though nevertheless $0 < m_0 < M_*$), there follows the dispersion relation

$$\varepsilon(\mathrm{p}) = \sqrt{c^2 p^2 + \varepsilon_0^2} \approx \varepsilon_0 + \frac{p^2}{2m_0} \quad , \quad p = |\vec{p}| \quad , \quad \varepsilon_0 = m_0 c^2 \; . \tag{7.4}$$

In the exponent of the exponential function $\exp[(\varepsilon - \mu)/k_B T]$ we can drop the rest energy $\varepsilon_0$, because both $\varepsilon = \varepsilon_0 + \cdots$ and $\mu = \varepsilon_0 + \cdots$ contribute the same amount of rest energy. We cannot drop, however, the rest energy $\varepsilon_0$ in the term

$$q(p) = 1 - \frac{\varepsilon(\mathrm{p})}{E_*} = 1 - \frac{\sqrt{c^2 p^2 + \varepsilon_0^2}}{E_*} = 1 - \frac{\varepsilon_0}{E_*}\sqrt{1 + \left(\frac{cp}{\varepsilon_0}\right)^2} \approx 1 - \frac{\varepsilon_0}{E_*} = q_0 \; , \tag{7.5}$$

occurring in the integrals (7.1) – (7.3). The now constant term $q = q_0$ changes these integrals into

$$P = \pm \frac{g_s}{q_0} \frac{4\pi}{h^3} k_B T \int_0^\infty p^2 \ln\left\{1 \pm q_0 \eta \exp\left[-\left(\frac{\varepsilon(p)}{k_B T}\right)\right]\right\} dp \; , \tag{7.6}$$

$$E = g_s \frac{4\pi}{h^3} V \int_0^\infty \frac{\varepsilon(p)}{\eta^{-1} \exp\left(\frac{\varepsilon(p)}{k_B T}\right) \pm q_0} p^2 dp \; , \tag{7.7}$$

$$N = g_s \frac{4\pi}{h^3} V \int_0^\infty \frac{1}{\eta^{-1} \exp\left(\frac{\varepsilon(p)}{k_B T}\right) \pm q_0} p^2 dp \; , \tag{7.8}$$

$$\varepsilon(\mathrm{p}) = \frac{p^2}{2m_0} \quad , \quad \eta = \exp(\beta \mu) \quad , \quad \beta = \frac{1}{k_B T} \quad , \quad q_0 = \left(1 - \frac{\varepsilon_0}{E_*}\right) . \tag{7.9}$$

Because of the small momenta $p \ll P_*$ in the non-relativistic case, the upper limit of the two integrals has been shifted to $P_* \to \infty$. The evaluation of the integrals (7.6) – (7.8) follows the procedure known in the statistics of ideal quantum gases (see [20], e.g.), with condition (6.15) to be taken into consideration in the Bose case.

Therefore, the exact solutions of (7.6) – (7.8) read

$$P = \frac{g_s}{q_0} \frac{k_B T}{\lambda^3} \begin{Bmatrix} f_{\frac{5}{2}}(q_0 \eta) \\ g_{\frac{5}{2}}(q_0 \eta) \end{Bmatrix} \quad , \tag{7.10}$$

$$E = \frac{g_s}{q_0} \frac{3}{2} \frac{V}{\lambda^3} k_B T \begin{Bmatrix} f_{\frac{5}{2}}(q_0 \eta) \\ g_{\frac{5}{2}}(q_0 \eta) \end{Bmatrix} = \frac{3}{2} P \cdot V \quad , \tag{7.11}$$

$$N = \frac{g_s}{q_0} \frac{V}{\lambda^3} \begin{Bmatrix} f_{\frac{3}{2}}(q_0 \eta) \\ g_{\frac{3}{2}}(q_0 \eta) \end{Bmatrix} \quad , \tag{7.12}$$

where $\lambda = h/\sqrt{2\pi m_0 k_B T}$ is the thermal de Broglie wavelength. The upper functions apply to fermions, the lower ones to bosons. $f_\nu(x)$ and $g_\nu(x)$ are the known generalized $\zeta$-functions of the type

$$f_\nu(x) = \sum_{r=1}^{\infty} (-1)^r \frac{x^r}{r^\nu} \quad , \quad g_\nu(x) = \sum_{r=1}^{\infty} \frac{x^r}{r^\nu} \quad . \tag{7.13}$$

### 7.1.1 Classical limit case

The classical limit case is characterized by $(\lambda^3/v) \ll 1$, i.e., by a high specific volume $v = V/N$ (low particle density) or high temperatures $(k_B T)$. With (7.12), this means $(q_0 \eta) \ll 1$, though. For this case, the Taylor expansions of the $f_\nu$ and $g_\nu$ functions read

$$f_\nu(x) = x - \frac{x^2}{2^\nu} + \cdots \quad , \quad g_\nu(x) = x + \frac{x^2}{2^\nu} + \cdots \quad , \quad x = q_0 \eta \ll 1 \quad . \tag{7.14}$$

Thus we obtain, in the usual way, the following quantum corrections of the thermal equation of state for fermions (upper sign) and bosons (lower sign)

$$P \cdot V = N k_B T \left( 1 \pm \frac{q_0}{g_s} \frac{(\lambda^3/v)}{2^{\frac{5}{2}}} \right) \quad , \quad q_0 = 1 - \frac{m_0}{M_*} \quad . \tag{7.15}$$

The associated energy can be read immediately from (7.11). Here we can readily see that, with $q_0 \to 0$ or $m_0 \to M_*$, respectively, all quantum effects disappear: According to (7.10) – (7.12) and with (7.14), a non-relativistic ideal quantum gas, the particles of which reach the size of Planck's mass, will for all temperature ranges always have the classical equation of state $PV = N k_B T$.

### 7.1.2 Limit case of high degeneration

The limit case of high degeneration is characterized by the condition $(\lambda^3/v) \approx 1$, i.e., by a low specific volume $v = V/N$ (high particle density) or low temperatures $(k_B T)$.

In the case of the f e r m i o n s and for temperatures below the Fermi temperature $T < T_F$, this will, with (7.10) – (7.12), lead to a constant pressure $P$, a constant energy $E$ and a constant chemical potential $\mu$,

$$P \cdot V = \frac{2}{5} N \varepsilon_F \quad , \quad E = \frac{3}{2} PV = \frac{3}{5} N \varepsilon_F \quad , \quad \mu = \varepsilon_F \quad , \tag{7.16}$$

$$\varepsilon_F = \frac{p_F^2}{2m_0} \quad , \quad p_F = \left(\frac{q_0}{g_s} 6\pi^2 \cdot \frac{N}{V}\right)^{\frac{1}{3}} \cdot \hbar \quad , \quad k_B T_F = \varepsilon_F \quad , \tag{7.17}$$

where $\varepsilon_F$ is the Fermi energy and $p_F$ the Fermi momentum.

In the case of b o s o n s, the requirement of $(\lambda^3/v) \approx 1$ for temperatures below the critical Einstein temperature $T < T_E$ leads to Bose-Einstein condensation. This happens if, at a specified density $N/V$ and with decreasing temperature $T$, the number $N_\varepsilon$ of particles occupying the excited states $\varepsilon \neq 0$ reaches its maximum level $N_\varepsilon^{\max}$. This level is given in (7.12) at $(q_0 \eta) \to 1$,

$$N_\varepsilon^{\max} = \frac{g_s}{q_0} \frac{V}{\lambda^3} g_{\frac{3}{2}}(1) \quad . \tag{7.18}$$

At a constant density $(N/V) = const.$, this defines the critical Einstein temperature $T = T_E$,

$$k_B T_E = 2\pi \left(\frac{q_0}{g_s \cdot g_{\frac{3}{2}}(1)} \frac{N}{V}\right)^{\frac{2}{3}} \cdot \frac{\hbar^2}{m_0} \quad , \quad g_{\frac{3}{2}}(1) = \zeta(\tfrac{3}{2}) = 2{,}612 \quad . \tag{7.19}$$

Below this temperature $T = T_E$, the rest $N_0 = N - N_\varepsilon^{\max}$ condenses into the ground state $\varepsilon = 0$. Taking the critical temperature $T_E$ from (7.19) into consideration, we calculate, on the one hand,

$$N_0 = N - N_\varepsilon^{\max} = N - \frac{g_s}{q_0} \frac{V}{\lambda^3} g_{\frac{3}{2}}(1) = N \left[1 - \left(\frac{T}{T_E}\right)^{\frac{3}{2}}\right] \quad , \quad T < T_E \quad . \tag{7.20}$$

On the other hand, according to (6.19), the number $N_0$ of particles in the ground state $\varepsilon = 0$ at $T \to 0$ (consider $\alpha = -(\mu/k_B T) \xrightarrow{T \to 0} 0$) has the very large, but f i n i t e value $N_0^{max}$:

$$N_0 = \frac{g_s}{\exp(\alpha) - q_0} \quad , \quad N_0^{max} = \lim_{T \to 0} N_0 = \frac{g_s}{1 - q_0} = \frac{g_s}{1 - \left(1 - \frac{\varepsilon_0}{E_*}\right)} = g_s \frac{M_*}{m_0} . \quad (7.21)$$

One can see that a Bose-Einstein condensate at $T \to 0$ should accommodate maximally $N_0^{max} \sim M_*/m_0$ particles of a Bose gas. The formula (7.20), then, is true only for the case that, with $T < T_E$ at the limit $T \to 0$, the condensate contains all $N$ particles of the Bose gas, and that, according to (7.21), the condition $N \leq N_0^{max}$ is satisfied. For an H-atom gas, e.g., $N_0^{max} = M_*/m_0 = M_{PL}/m_p = 1.30 \cdot 10^{19}$, if $M_* = M_{Pl}$ is the Planck mass and $m_0 = m_p$ is the proton mass. These are particle numbers occurring, say, in a volume of 1 cm³ of an ideal gas at room temperature and normal pressure. However, the absolute particle numbers nowadays achievable in a Bose condensate are only around $10^8$ particles per condensate [21], [22]. What is remarkable about the result (7.21) is that, *in principle,* maximally $N = N_0^{max}$ particles may be contained in a condensate. If the condensate were just filled with $N_0^{max}$ particles of the gas, it would behave like a single droplet (elementary particle) with a Planck rest mass $M_* = N_0^{max} \cdot m_0$. The addition of further particles should then lead to a second and perhaps further, spatially separated condensate regions. In this way, an ideal Planck mass gas from many individual Bose condensates could form. This presumption is consistent in the sense of the hypothesis (1.2): If there is a maximum quantum leap $\Delta E \leq E_*$ in nature, there should only exist such Bose-Einstein condensates that, individually, do not exceed a rest energy $E \leq E_* = N_0^{max} \cdot m_0 c^2 = M_* c^2$.

If it could be experimentally verified that the condensation rate $N_0$ of a Bose-Einstein condensate at temperatures $T < T_E$ cannot be greater than $N_0 \leq N_0^{max} = g_s M_*/m_0$, one would have a direct proof of the existence and size of the limit mass $M_* = \alpha \cdot M_{Pl}$ and also of the limit energy $E_* = M_* c^2$.

### 7.1.3 Limit cases of the $q_0$ term

What is new in the results of chapter 7.1 is shown by the appearance of the constant term $q_0 = (1 - \varepsilon_0/E_*) = (1 - m_0/M_*)$ in the quantum corrections for the classical limit case (7.15), as well as in the case of degenerations (7.17) and (7.19) of Fermi and Bose gases. However, even in the distribution formulas (6.9) and (6.20) it is possible to manipulate the kind of statistics with the $q_0$ parameter.

To make this more obvious, let us introduce the following abbreviations:

$$x = \frac{\varepsilon}{E_*} \quad , \quad y = \frac{\mu}{E_*} \quad , \quad \theta = \frac{k_B T}{E_*} \quad . \tag{7.22}$$

With the conditions (7.4) for non-relativistic quantum gases and

$$\frac{\varepsilon}{k_B T} = \frac{x}{\theta} \quad , \quad \frac{\mu}{k_B T} = \frac{y}{\theta} \quad , \quad x_0 = \frac{\varepsilon_0}{E_*} = \frac{m_0}{M_*} \quad , \quad q_0 = 1 - \frac{m_0}{M_*} = 1 - x_0 \tag{7.23}$$

the distribution formulas (6.9) and (6.20) with $\varepsilon_k \to \varepsilon(p)$ now read

$$\bar{n}(x, y, \theta) = \frac{1}{\exp\left(\frac{x-y}{\theta}\right) - (1 - x_0)} \quad \text{bosons} \; , \tag{7.24}$$

$$\bar{n}(x, y, \theta) = \frac{1}{\exp\left(\frac{x-y}{\theta}\right) + (1 - x_0)} \quad \text{fermions} \; . \tag{7.25}$$

Typical function behaviours for the two cases $\varepsilon_0 \ll E_*$ ($x_0 \ll 1$) (common quantum statistics) and $\varepsilon_0 \to E_*$ ($x_0 \to 1$) (new quantum statistics) are shown in Figs. 12 and 13 in the appendix.

$\varepsilon_0 \ll E_*$ ($x_0 \ll 1$):

Let us first look at the limit case $\varepsilon_0 \ll E_*$ or $q_0 = 1 - x_0 = 1 - (\varepsilon_0/E_*) \approx 1$. In this case, there appears in (7.6) – (7.8) the common quantum statistics of ideal gases in their presently known type for fermions and bosons. The quantum corrections for the classical limit case (7.15) and the statements in the case of high degeneration (7.16) – (7.19) are essentially maintained because of $q_0 \approx 1$. A closer look, however, will reveal some remarkable new properties for the low-temperature limit case.

For f e r m i o n s, according to (6.8), the number $N_0$ of particles in the ground state (with $T \to 0$, $\mu \to \varepsilon_F$) has the value

$$N_0^{\max} = N_0(T=0) = \frac{g_s}{0 + q_0} = \frac{g_s}{1 - x_0} \quad , \quad \bar{n}_0^{\max} = \frac{N_0^{\max}}{g_s} = \frac{1}{1 - \frac{m_0}{M_*}} \approx 1 + \frac{m_0}{M_*} \; , \tag{7.26}$$

so that the mean occupation number $\bar{n}_0$ of the ground state no longer has exactly the value $\bar{n}_0^{\max} = 1$ as one would expect for fermionic gases according to Pauli's principle. For the ³He-gas, e.g., one finds, with $M_* = M_{Pl}$, $m_0 = 2m_p + m_n \approx 3m_p$ a value of $m_0/M_* \approx 2{,}31 \cdot 10^{-19}$.

If we increase the rest mass $m_0$ of the gas particles up to Planck's mass $M_*$, the mean occupation number $\bar{n}_0^{max}$ of the ground state can grow immensely, and the result is: with $m_0 \rightarrow M_*$, the Fermi gas acquires increasingly bosonic properties.

For b o s o n s, according to (6.19), the number $N_0$ of particles in the ground state (with $T \rightarrow 0$, $\alpha \rightarrow 0$) has the value

$$N_0^{max} = N_0(T=0) = \frac{g_s}{1-q_0} = \frac{g_s}{x_0} \quad , \quad \bar{n}_0^{max} = \frac{N_0^{max}}{g_s} = \frac{M_*}{m_0} \quad , \tag{7.27}$$

so that the mean occupation number $\bar{n}_0^{max}$ of the ground state no longer has the infinitely high value $\bar{n}_0^{max} \rightarrow \infty$ as known with bosonic gases. However, if we increase the rest mass $m_0$ of the gas particles up to Planck's mass $M_*$, the mean occupation number $\bar{n}_0^{max}$ of the ground state can decrease down to $\bar{n}_0^{max} = 1$, and the result is: with $m_0 \rightarrow M_*$, the Bose gas increasingly adopts fermionic properties.

$\varepsilon_0 \approx E_* (x_0 \approx 1)$ :

Our quantum statistics undergoes an interesting alteration in the extreme case of $\varepsilon_0 \approx E_*$ or $q_0 = 1 - x_0 = 1 - (\varepsilon_0 / E_*) \approx 0$. In non-relativistic quantum statistics we achieve $q_0 = (1-(m_0/M_*)) \rightarrow 0$ by an increasing rest mass of the particles, $m_0 \rightarrow M_*$, with gas particle speeds kept low. As can be seen already in the distribution formulas (7.24) and (7.25), the entire quantum statistics of ideal gases changes, for $q_0 \rightarrow 0$, into the Boltzmann statistics of an ideal gas then consisting of Planck masses. As a consequence, with $q_0 \rightarrow 0$, all quantum corrections of the classical thermal equation of state (7.15) completely disappear. In the case of low-temperature degeneration, too, with $q_0 \rightarrow 0$, Fermi momentum (7.17) and critical temperature (7.19) for Fermi respectively Bose gas decline towards zero. The upshot is that with $q_0 \rightarrow 0$, i.e., with the increase in the particles' rest mass $m_0 \rightarrow M_*$ alone, all quantum effects of the ideal quantum gases are suppressed. In passing to the limit $q_0 \rightarrow 0$ (irrespective of the spin character), the ideal, non-relativistic quantum gas turns a classical ideal Boltzmann gas!

For the time being, this applies to a temperature range in which the mean particle energy $\varepsilon = m_0 v^2 / 2$ is low compared to the rest energy $\varepsilon_0 = m_0 c^2$ of the gas particles:

$$\frac{p^2}{2m_0} \ll m_0 c^2 \quad , \quad \frac{p^2}{2m_0} \approx \pi \cdot k_B T \quad , \quad \pi \cdot k_B T \ll \varepsilon_0 \quad . \tag{7.28}$$

Expressed in de-Broglie wavelengths, we obtain, with (7.28), the relations

$$\lambda_{th} = \frac{h}{p} = \frac{h}{\sqrt{2\pi m_0 k_B T}} \quad , \quad \lambda_c = \frac{h}{m_0 c} \quad , \quad \lambda_{th} \gg \lambda_c \quad . \tag{7.29}$$

If the thermal wavelength $\lambda_{th}$ is long compared to the Compton wavelength $\lambda_c$ of the gas particles, the gas shows a non-relativistic behaviour. For passing to the limit $q_0 \to 0$, i.e. $m_0 \to M_*$, however, this means that a gas with Planck mass particles shows a non-relativistic Boltzmann behaviour even for relatively high temperatures $k_B T \ll M_* c^2 = E_*$ or $\lambda_{th} \gg \hbar/M_* c = L_*$ ($L_*$ Planck length). In other words: An ideal Planck mass gas only shows deviations from the ideal nonrelativistic Boltzmann gas if its thermal wavelength comes close to the Planck length.

The new properties of ideal quantum gases with ultra-relativistic particle speeds are described in the following chapter.

**7.2 Ultra-relativistic quantum gases**

In den exact equations of state (7.1) – (7.3), we now use the approximate dispersion relation for sufficiently high particle momenta $p \gg m_0 c$, so that we can work with

$$\varepsilon(p) = \sqrt{c^2 p^2 + \varepsilon_0^2} \approx c p \sqrt{1 + \left(\frac{m_0 c}{p}\right)^2} \approx c p \tag{7.30}$$

Varying with their spin properties, such gas particles then behave, in a good approximation, like zero-rest-mass photons or neutrinos. A consequence of this is that the chemical potential $\mu$ vanishes and the grand canonical potential $\Phi$ becomes identical with the free energy $F$:

$$\Phi = U - TS - \mu N = F = U - TS = -PV \quad . \tag{7.31}$$

If, in the integrals (7.1) – (7.3), we perform the substitution $cp = \varepsilon$ (with $cP_* = E_*$), we get (upper sign: fermions; lower sign: bosons):

$$P = \pm g_s \frac{4\pi}{(hc)^3} k_B T \int_0^{E_*} \frac{\varepsilon^2}{\left(1 - \frac{\varepsilon}{E_*}\right)} \ln\left\{1 \pm \left(1 - \frac{\varepsilon}{E_*}\right) \exp\left[-\left(\frac{\varepsilon}{k_B T}\right)\right]\right\} d\varepsilon \tag{7.32}$$

$$E = g_s \frac{4\pi}{(hc)^3} V \int_0^{E_*} \frac{\varepsilon^3}{\exp\left(\frac{\varepsilon}{k_B T}\right) \pm \left(1 - \frac{\varepsilon}{E_*}\right)} d\varepsilon \tag{7.33}$$

$$N = g_s \frac{4\pi}{(hc)^3} V \int_0^{E_*} \frac{\varepsilon^2}{\exp\left(\dfrac{\varepsilon}{k_B T}\right) \pm \left(1 - \dfrac{\varepsilon}{E_*}\right)} d\varepsilon \qquad (7.34)$$

In this case, the upper integration limit $E_*$ needs to be kept exactly. The term $q = (1 - \varepsilon/E_*)$ in (7.32), too, can no longer be placed in front of the integral sign.

### 7.2.1 Zero-rest-mass fermions (neutrinos)

In den equations of state (7.32) – (7.34), the upper sign applies. We first investigate the new curve of the spectral energy density $\bar{u}$, the form of which can be read from (7.33):

$$E = V \int_0^{\Omega_*} \bar{u}(\omega, T)\, d\omega = V \int_0^{E_*} \left(\frac{\bar{u}}{\hbar}\right) d\varepsilon \quad , \quad \varepsilon = \hbar\omega \, , \qquad (7.35)$$

$$\bar{u} = \left(\frac{g_s 4\pi}{(hc)^3} \hbar \varepsilon^2\right) \bar{\varepsilon} = \frac{g_s \omega^2}{2\pi^2 c^3} \cdot \bar{\varepsilon} \, ,$$

$$\bar{\varepsilon} = \frac{\varepsilon}{\exp\left(\dfrac{\varepsilon}{k_B T}\right) + \left(1 - \dfrac{\varepsilon}{E_*}\right)} \quad , \quad \bar{n} = \frac{1}{\exp\left(\dfrac{\varepsilon}{k_B T}\right) + \left(1 - \dfrac{\varepsilon}{E_*}\right)} \, . \qquad (7.36)$$

Here, $\bar{\varepsilon}$ is the mean energy, which has the energy level $\varepsilon = \hbar\omega$ at the temperature $T$, and $\bar{n}$ is its mean occupation number.

Let us state several limit cases of the new fermionic distribution $\bar{\varepsilon}(T, \omega)$:

for $E_* \to \infty$ (with any $\omega, T$) there appears the Fermi distribution $\bar{\varepsilon} = \bar{\varepsilon}_{FD}$,

for $\varepsilon \to E_*$ (with any $T$) there appears the Boltzmann distribution ,

for $T \to \infty$ (with any $\varepsilon < E_*$) there appears $\bar{\varepsilon} = E_*/(2\frac{E_*}{\varepsilon} - 1)$, so that at $\varepsilon \to E_*$ the upper limit $\bar{\varepsilon} = E_*$ is attained.

Other ways of exact notation for the mean energy are

$$\bar{\varepsilon} = \frac{\bar{\varepsilon}_{FD}}{1 - \dfrac{\bar{\varepsilon}_{FD}}{E_*}} \quad , \quad \frac{1}{\bar{\varepsilon}} = \frac{1}{\bar{\varepsilon}_{FD}} - \frac{1}{E_*} \quad , \quad \bar{\varepsilon}_{FD} = \frac{\hbar\omega}{\exp\left(\dfrac{\hbar\omega}{k_B T}\right) + 1} \, , \qquad (7.37)$$

with $\bar{\varepsilon}_{FD}$ being the well-known Fermi-Dirac distribution.

Let us investigate the curve of the spectral energy density $\bar{u}(\omega,T)$ by introducing the dimensionless variables $z$ and $\theta$ into (7.36):

$$z = \frac{\varepsilon}{k_B T} \quad , \quad \theta = \frac{k_B T}{E_*} \quad , \quad \theta z = \frac{\varepsilon}{E_*} \quad , \tag{7.38}$$

$$\bar{u}(z,\theta) = \frac{g_s (k_B T)^3}{2\pi^2 \hbar^2 c^3} \left\{ \frac{z^3}{\exp(z) + (1-\theta z)} \right\} ,$$

$$\bar{\varepsilon} = k_B T \left\{ \frac{z}{\exp(z) + (1-\theta z)} \right\} \quad , \quad \bar{n} = \left\{ \frac{1}{\exp(z) + (1-\theta z)} \right\} . \tag{7.39}$$

The function behaviour of $\bar{u}(z,\theta)$ is shown in Fig. 3 in the appendix. The maximum $z = z_0$ of the function $\bar{u} = \bar{u}(z,\theta)$ with $\partial \bar{u}/\partial z = 0$ results from the zero of the function

$$Q(z,\theta) = (3-z)\exp(z) + (3-2\theta z) \quad , \quad Q(z_0,\theta) = 0 , \tag{7.40}$$

with the term containing the parameter $\theta$ indicating the change relative to the known Fermi-Dirac distribution with $\theta = 0$. The following limit cases can be evaluated simply:

$$\begin{aligned}
z_0 &= z_{FD} = 3.131 & \theta &= 0 \quad (E_* \to \infty) , \\
z_0 &\approx z_{FD}(1 - 0.24 \cdot \theta) & \theta &\ll 1 \quad (k_B T \ll E_*) , \\
z_0 &= 3 & \theta &= 1/2 \quad (k_B T = E_*/2), \\
z_0 &\approx 3/\theta & \theta &\gg 1 \quad (k_B T \gg E_*) .
\end{aligned} \tag{7.41}$$

Generally, the law of shifting the distribution maximum now reads

$$(\hbar \omega)_{max} = z_0(\theta) \cdot (k_B T) . \tag{7.42}$$

Due to the temperature dependence of $z_0(\theta)$ that can be calculated from (7.40), the maximum, with increasing temperature, shifts more slowly towards increasing frequencies compared to Wien's known (fermionic) displacement law. Some examples are given in (7.41) and graphically in Fig. 4 (see appendix). This shift, though, is significant only for temperatures near the Planck temperature $k_B T = E_*$. If, e.g., the temperature $k_B T = E_*/100$, there follows $z_0 = 0.9976 \cdot z_{FD} = 3,123$, i.e., a decrease in the $z_0$ value by 0.008 only.

To be able to display also the function behaviours of the most important thermodynamic functions $P, E, N$, we introduce the parameter $\theta = k_B T / E_*$ in (7.32) – (7.34) by substitution of (7.38) and transcribe the factors preceding the integrals to the Planck quantities defined in chapter 2. For the ultrarelativistic fermionic gas, there result the functions

$$PV = g_s N_* E_* \theta^4 \int_0^{\frac{1}{\theta}} \frac{z^2}{1-\theta z} \ln[1+(1-\theta z)\exp(-z)]\, dz \;, \tag{7.43}$$

$$U = g_s N_* E_* \theta^4 \int_0^{\frac{1}{\theta}} \frac{z^3\, dz}{\exp(z)+(1-\theta z)} \;, \tag{7.44}$$

$$N = g_s N_* \theta^3 \int_0^{\frac{1}{\theta}} \frac{z^2\, dz}{\exp(z)+(1-\theta z)} \;, \tag{7.45}$$

which are shown in Figs. 5, 7 and 8 in the appendix.

Thus, at the same time, we know the free energy $F = -PV$, and the entropy $S$ ( Fig. 9) is determined from

$$S = -\left(\frac{\partial F}{\partial T}\right)_V = g_s k_B N_* \frac{\partial}{\partial \theta}\left(\frac{PV}{g_s k_B N_*}\right). \tag{7.46}$$

For the ratio $w(\theta) = PV/U$, there appears the function behaviour shown in Fig. 11, with $\lim_{\theta \to 0} w(\theta) = 1/3$ and the asymptote $\lim_{\theta \to \infty} w(\theta) = 4.72\, C_\infty^F \cdot \theta + (3.40\, C_\infty^F - 1)$.

If we set $\theta = 0$ ($E_* \to \infty$) in the integrals in (7.43) – (7.45), there appear the known Fermi-Dirac functions (e.g., for the neutrino radiation with $g_s = 1$),

$$PV = \frac{1}{3} U \tag{7.47}$$

$$U = g_s N_* E_* \theta^4 \int_0^\infty \frac{z^3\, dz}{\exp(z)+1} = g_s N_* E_* \theta^4 \left(\frac{7}{8}\right)\Gamma(4)\zeta(4) = \frac{4}{c}\sigma_F V T^4 \;, \tag{7.48}$$

$$N = g_s N_* \theta^3 \int_0^\infty \frac{z^2\, dz}{\exp(z)+1} = g_s N_* \theta^3 \left(\frac{3}{4}\right)\Gamma(3)\zeta(3) = \kappa_F V T^3 \;, \tag{7.49}$$

$$F = -PV = -\frac{1}{3} U \;, \quad S = -\left(\frac{\partial F}{\partial T}\right)_V = \frac{4}{3}\frac{U}{T} \;, \tag{7.50}$$

$$\sigma_F = \frac{7}{8}\sigma_B \;, \quad \kappa_F = \frac{3}{4}\kappa_B \;, \tag{7.51}$$

where $\sigma_B$ and $\kappa_B$ are the corresponding constants of the bosons (cf. (7.80)).

It is remarkable that, for $k_B T \gg E_*$ ($\theta \to \infty$), some thermodynamic potentials have finite limit values now. We state the respective values in the formulas below. Substituting $\theta z = x$ in (7.43) – (7.45) and then setting $\theta \to \infty$, we obtain

$$\lim_{\theta \to \infty} U = U_\infty = g_s N_* E_* \int_0^1 \frac{x^3}{2-x} dx = \left(8\ln 2 - \frac{16}{3}\right) g_s N_* E_* \approx 0.212\, g_s N_* E_* \,, \quad (7.52)$$

$$\lim_{\theta \to \infty} N = N_\infty = g_s N_* \int_0^1 \frac{x^2}{2-x} dx = \left(4\ln 2 - \frac{5}{2}\right) g_s N_* \approx 0.273\, g_s N_* \,, \quad (7.53)$$

$$\lim_{\theta \to \infty} PV = -\lim_{\theta \to \infty} F \to g_s N_* E_* \theta \int_0^1 \frac{x^2}{1-x} \ln(2-x)\, dx = g_s N_* E_* C_\infty^F \cdot \theta \,, \quad (7.54)$$

$$C_\infty^F = -4\ln 2 + \frac{\pi^2}{12} + \frac{9}{4} \approx 0.300 \,. \quad (7.55)$$

With (7.54), (7.46) as well the definition of Planck density $u_*$ and Planck particle number $N_*$ from chapter 2, we get then

$$\lim_{\theta \to \infty} P = \lim_{\theta \to \infty}\left(-\frac{F}{V}\right) \to g_s C_\infty^F \cdot u_* \cdot \theta \,, \quad (7.56)$$

$$\lim_{\theta \to \infty} S = S_\infty = -\lim_{\theta \to \infty} \frac{\partial F}{\partial T} = \frac{\partial}{\partial T}\left(-\lim_{\theta \to \infty} F\right) = g_s C_\infty^F N_* k_B = \tilde{N}_*^F \cdot k_B \,, \quad (7.57)$$

in which we have summarily set $\tilde{N}_*^F = g_s C_\infty^F N_*$.

The energy (7.52) or entropy (7.57) in this limit case $\theta \to \infty$ ($T \to \infty$) is about equal to the number $N_*$ of Planck volumes $V_*$ accommodated in a specified volume $V$ ($N_* = V/V_*$), multiplied by the Planck energy $E_*$ or the Boltzmann constant $k_B$, respectively. The number of Neutrinos (7.53), too, is about equal to the number of Planck volumes $V_*$ filling the specified volume $V$.

The equation of state of such a matter configuration obviously is that of an ideal, non-interacting gas of $\tilde{N}_*^F$ Planck masses at extremely high temperatures $T$, as can be read from (7.52) – (7.57):

$$PV = -F \xrightarrow{T \to \infty} -(U_\infty - T S_\infty) \approx T S_\infty \,, \quad (7.58)$$

$$PV \approx \tilde{N}_*^F k_B T \,. \quad (7.59)$$

We note in addition that in the limit case $T \to \infty$, by means of Boltzmann's entropy equation $S = k_B \ln \Omega$, we can also state a maximum number of states $\Omega_\infty$. Since

$$S_\infty = k_B \ln \Omega_\infty = k_B g_s C_\infty^F N_* = k_B \tilde{N}_*^F \tag{7.60}$$

is the fermion entropy maximally possible in the volume $V$, maximally $\Omega_\infty$ states of one fermion kind can be attained in this volume,

$$\Omega_\infty = \exp\left(\frac{S_\infty}{k_B}\right) = \left[\exp\left(g_s C_\infty^F\right)\right]^{N_*} . \tag{7.61}$$

For neutrinos of a kind with $g_s = 1$ and $C_\infty^F = 0.30$, we can compute the maximum entropy in $V$, which is $S_\infty^{(1)} = C_\infty^F N_* k_B = 0.30 \, N_* k_B$. Hence, the maximally possible number of states in $V$ is $\Omega_\infty^{(1)} = [\exp(0.30)]^{N_*} = [1.35]^{N_*}$. With three kinds of neutrinos, the maximum entropy in $V$ triples to

$$S_\infty^{(3)} = 3 \cdot S_\infty^{(1)} = 3 \cdot C_\infty^F N_* k_B = 0.90 \, N_* k_B \tag{7.62}$$

resulting in a maximum number of states in $V$ of

$$\Omega_\infty^{(3)} = [\exp(0.90)]^{N_*} = [2.46]^{N_*} . \tag{7.63}$$

### 7.2.2 Zero-rest-mass bosons (photons)

In the equations of state (7.32) – (7.34), the lower sign applies. Let us first examine the new curve of the spectral energy density $\bar{u}$, the form of which can be read from (7.33):

$$E = V \int_0^{\Omega_*} \bar{u}(\omega, T) \, d\omega = V \int_0^{E_*} \left(\frac{\bar{u}}{\hbar}\right) d\varepsilon \quad , \quad \varepsilon = \hbar \omega , \tag{7.64}$$

$$\bar{u} = \left(\frac{g_s 4\pi}{(hc)^3} \hbar \varepsilon^2\right) \bar{\varepsilon} = \frac{g_s \omega^2}{2\pi^2 c^3} \cdot \bar{\varepsilon} \quad ,$$

$$\bar{\varepsilon} = \frac{\varepsilon}{\exp\left(\frac{\varepsilon}{k_B T}\right) - \left(1 - \frac{\varepsilon}{E_*}\right)} \quad , \quad \bar{n} = \frac{1}{\exp\left(\frac{\varepsilon}{k_B T}\right) - \left(1 - \frac{\varepsilon}{E_*}\right)} . \tag{7.65}$$

where $\bar{\varepsilon}$ is the mean energy of the energy level $\varepsilon = \hbar \omega$ at the temperature $T$, and $\bar{n}$ its mean occupation number.

Here are some limit cases of the new bosonic distribution $\bar{\varepsilon}(T,\omega)$:

For $E_* \to \infty$ (with any $\omega, T$), there appears the Bose distribution $\bar{\varepsilon} = \bar{\varepsilon}_{BE}$,

for $\varepsilon \to E_*$ (with any $T$) there appears the Boltzmann distribution, and for $T \to \infty$ (with any $\varepsilon < E_*$) there appears $\bar{\varepsilon} = E_*$.

Other exact notations for the mean energy $\bar{\varepsilon}$ are

$$\bar{\varepsilon} = \frac{\bar{\varepsilon}_{BE}}{1+\frac{\bar{\varepsilon}_{BE}}{E_*}} \quad , \quad \frac{1}{\bar{\varepsilon}} = \frac{1}{\bar{\varepsilon}_{BE}} + \frac{1}{E_*} \quad , \quad \bar{\varepsilon}_{BE} = \frac{\hbar\omega}{\exp\left(\frac{\hbar\omega}{k_BT}\right)-1} \quad , \tag{7.66}$$

where $\bar{\varepsilon}_{BE}$ is the well-known Bose-Einstein distribution.

To investigate the curve of the spectral energy density $\bar{u}(\omega,T)$, we introduce the dimensionless variables $z$ and $\theta$ into (7.65):

$$z = \frac{\varepsilon}{k_BT} \quad , \quad \theta = \frac{k_BT}{E_*} \quad , \quad \theta z = \frac{\varepsilon}{E_*} \quad , \tag{7.67}$$

$$\bar{u}(z,\theta) = \frac{g_s(k_BT)^3}{2\pi^2\hbar^2c^3}\left\{\frac{z^3}{\exp(z)-(1-\theta z)}\right\} \quad ,$$

$$\bar{\varepsilon} = k_BT\left\{\frac{z}{\exp(z)-(1-\theta z)}\right\} \quad , \quad \bar{n} = \left\{\frac{1}{\exp(z)-(1-\theta z)}\right\} . \tag{7.68}$$

The function behaviour of $\bar{u}(z,\theta)$ is shown in Fig. 2 (see Appendix). The maximum $z = z_0$ of the function $\bar{u} = \bar{u}(z,\theta)$ with $\partial\bar{u}/\partial z = 0$ results from the zero of the function

$$Q(z,\theta) = (3-z)\exp(z) - (3-2\theta z) \quad , \quad Q(z_0,\theta) = 0 \quad , \tag{7.69}$$

in which the term containing the parameter $\theta$ indicates the change relative to the known Bose Einstein distribution with $\theta = 0$. The following limit cases can be evaluated simply:

$$\begin{aligned}
z_0 = z_{BE} = 2.821 & \qquad \theta = 0 \quad (E_* \to \infty) \;, \\
z_0 \approx z_{BE}(1+0.41\cdot\theta) & \qquad \theta \ll 1 \quad (k_BT \ll E_*) \;, \\
z_0 = 3 & \qquad \theta = 1/2 \quad (k_BT = E_*/2) \;, \\
z_0 \approx \ln(2\theta) & \qquad \theta \gg 1 \quad (k_BT \gg E_*) \;.
\end{aligned} \tag{7.70}$$

Generally, the law for shifting the distribution maximum now reads

$$(\hbar\omega)_{max} = z_0(\theta)\cdot(k_BT) \;. \tag{7.71}$$

Due to the temperature dependence of $z_0(\theta)$ that can be calculated from (7.69), the maximum, with increasing temperature, shifts more rapidly towards increasing frequencies compared to Wien's known (bosonic) displacement law. Some examples are given in (7.70) and graphically in Fig.4 (see Appendix). This shift, though, is significant only for temperatures near the Planck temperature $k_B T = E_*$. If, e.g., the temperature $k_B T = E_*/100$, there follows $z_0 = 1.0041 \cdot z_{BE} = 2.833$, i.e., an increase of the $z_0$ value by $0.012$ only.

To be able to display also the function behaviours of the most important thermodynamic functions $P, E, N$, we introduce the parameter $\theta = k_B T / E_*$ in (7.32) – (7.34) by substitution of (7.67) and transcribe the factors preceding the integrals to the Planck quantities defined in chapter 2. For the ultrarelativistic Bose gas, there result the functions

$$PV = -g_s N_* E_* \theta^4 \int_0^{\frac{1}{\theta}} \frac{z^2}{1-\theta z} \ln[1-(1-\theta z)\exp(-z)]\, dz \;, \tag{7.72}$$

$$U = g_s N_* E_* \theta^4 \int_0^{\frac{1}{\theta}} \frac{z^3\, dz}{\exp(z)-(1-\theta z)} \;, \tag{7.73}$$

$$N = g_s N_* \theta^3 \int_0^{\frac{1}{\theta}} \frac{z^2\, dz}{\exp(z)-(1-\theta z)} \;, \tag{7.74}$$

which are shown in Figs. 5, 6 and 8 in the appendix.

Thus, at the same time, we know the free energy $F = -PV$, and the entropy $S$ ( Fig. 9) is determined from

$$S = -\left(\frac{\partial F}{\partial T}\right)_V = g_s k_B N_* \frac{\partial}{\partial \theta}\left(\frac{PV}{g_s k_B N_*}\right) . \tag{7.75}$$

For the relation $w(\theta) = PV/U$ there appears the function behaviour shown in Fig. 10, with $\lim_{\theta \to 0} w(\theta) = 1/3$ and the asymptote $\lim_{\theta \to \infty} w(\theta) = 3C_\infty^B \cdot \theta + (3C_\infty^B - 1)$.

If we set $\theta = 0$ ($E_* \to \infty$) in the integrals in (7.72) – (7.74), there appear the known Bose-Einstein functions (e.g., for the photon radiation with $g_s = 2$ ),

$$PV = \frac{1}{3}U \tag{7.76}$$

$$U = g_s N_* E_* \theta^4 \int_0^\infty \frac{z^3\, dz}{\exp(z)-1} = g_s N_* E_* \theta^4 \cdot \Gamma(4)\zeta(4) = \frac{4}{c}\sigma_B V T^4 \;, \tag{7.77}$$

$$N = g_s N_* \theta^3 \int_0^\infty \frac{z^2\, dz}{\exp(z)-1} = g_s N_* \theta^3 \cdot \Gamma(3)\zeta(3) = \kappa_B V T^3 \;,\tag{7.78}$$

$$F = -PV = -\frac{1}{3}U \;,\quad S = -\left(\frac{\partial F}{\partial T}\right)_V = \frac{4}{3}\frac{U}{T} \;,\tag{7.79}$$

$$\sigma_B = \frac{g_s}{2}\frac{k^4 \pi^2}{60 c^2 \hbar^3} \;,\quad \kappa_B = g_s \frac{\zeta(3)}{\pi^2}\left(\frac{k}{\hbar c}\right)^3 ,\tag{7.80}$$

where $\sigma_B$ is the Stefan-Boltzmann constant.

It is remarkable that, for $k_B T \gg E_*$ ($\theta \to \infty$), some thermodynamic potentials have finite limit values now. We state the respective values in the formulas below. Substituting $\theta z = x$ in (7.72) – (7.74) and then setting $\theta \to \infty$, we obtain

$$\lim_{\theta \to \infty} U = U_\infty = g_s N_* E_* \int_0^1 x^2\, dx = \frac{1}{3} g_s N_* E_* \tag{7.81}$$

$$\lim_{\theta \to \infty} N = N_\infty = g_s N_* \int_0^1 x\, dx = \frac{1}{2} g_s N_* \tag{7.82}$$

$$\lim_{\theta \to \infty} PV = -\lim_{\theta \to \infty} F \to -g_s N_* E_* \theta \int_0^1 \frac{x^2}{1-x}\ln(x)\, dx = g_s N_* E_* \cdot C_\infty^B \cdot \theta \;,\tag{7.83}$$

$$C_\infty^B = \frac{\pi^2}{6} - \frac{5}{4} \approx 0.359 \;.\tag{7.84}$$

With (7.83), (7.75) as well the definition of Planck density $u_*$ and Planck particle number $N_*$ from chapter 2, we get then

$$\lim_{\theta \to \infty} P = \lim_{\theta \to \infty}\left(-\frac{F}{V}\right) \to g_s C_\infty^B \cdot u_* \cdot \theta \;,\tag{7.85}$$

$$\lim_{\theta \to \infty} S = S_\infty = -\lim_{\theta \to \infty}\frac{\partial F}{\partial T} = \frac{\partial}{\partial T}\left(-\lim_{\theta \to \infty} F\right) = g_s C_\infty^B N_* k_B = \tilde{N}_*^B \cdot k_B \;,\tag{7.86}$$

in which we have summarily set $\tilde{N}_*^B = g_s C_\infty^B N_*$.

The energy (7.81) or entropy (7.86) in this limit case $\theta \to \infty$ ($T \to \infty$) is about equal to the number $N_*$ of Planck volumes $V_*$ accommodated in a specified volume $V$ ($N_* = V/V_*$), multiplied by the Planck energy $E_*$ or the Boltzmann constant $k_B$, respectively.

The number of photons (7.82), too, is about equal to the number of Planck volumes $V_*$ filling the specified volume $V$.

The equation of state of such a matter configuration obviously is that of an ideal, non-interacting gas of $\tilde{N}_*^B$ Planck masses at extremely high temperatures $T$, as can be read from (7.81) – (7.86):

$$PV = -F \xrightarrow[T \to \infty]{} -(U_\infty - T S_\infty) \approx T S_\infty, \tag{7.87}$$

$$PV \approx \tilde{N}_*^B k_B T. \tag{7.88}$$

We note in addition that in the limit case $T \to \infty$, by means of Boltzmann's entropy equation $S = k_B \ln \Omega$, we can also state a maximum number of states $\Omega_\infty$. Since

$$S_\infty = k_B \ln \Omega_\infty = k_B g_s C_\infty^B N_* = k_B \tilde{N}_*^B \tag{7.89}$$

is the boson entropy maximally possible in the volume $V$, maximally $\Omega_\infty$ states of one boson kind can be attained in this volume:

$$\Omega_\infty = \exp\left(\frac{S_\infty}{k_B}\right) = \left[\exp\left(g_s C_\infty^B\right)\right]^{N_*}. \tag{7.90}$$

For photons with $g_s = 2$ and $C_\infty^B = 0.395$, we can compute the maximum entropy in $V$, which is $S_\infty = 2 C_\infty^B N_* k_B = 0.79 N_* k_B$. Hence, the maximally possible number of states in $V$ is

$$\Omega_\infty = [\exp(0.79)]^{N_*} = [2.20]^{N_*}. \tag{7.91}$$

## 8 Thermodynamics and conditional probability

We want to examine how the entropy S of the ideal quantum gases in case of energetically universally limited quantum leaps relates to certain probabilities. Let us look at the schematic representation of states in Fig. 1. We assume that each bundle of states $|k\rangle$ with the associated energy $\varepsilon_k$ and the $N_k$ quantum particles contained in it is a stand-alone thermodynamic system independent of the other bundles. The entropy $S_k$ is, then, just that from (2.4) with the $q_k$ factor. This factor assigns to each of the $G_k$ states in $|k\rangle$ an elementary probability $q_k$, which is equal for all these states. Therefore, with an occupation number of $N_k$ particles in $|k\rangle$, the entropy of the $|k\rangle$- bundle is

$$S_k = k_B \ln \Omega_k = k_B \ln \left[ \frac{G_k!}{\prod_{r=0}^{r_k} g_k^{(r)}!} \cdot q_k^{N_k} \right]. \tag{8.1}$$

With Stirling's formula and several transformations (cf. (3.7), (3.8)) we first get

$$\frac{S_k}{k_B} = G_k \ln G_k - \sum_{r=0}^{r_k} g_k^{(r)} \ln g_k^{(r)} + N_k \ln q_k \tag{8.2}$$

and, with (3.2) and (3.4),

$$\frac{S_k}{k_B} = G_k \left\{ \ln G_k - \sum_{r=0}^{r_k} \frac{g_k^{(r)}}{G_k} \ln g_k^{(r)} + \sum_{r=0}^{r_k} r \frac{g_k^{(r)}}{G_k} \ln q_k \right\}, \tag{8.3}$$

$$\frac{S_k}{k_B} = -G_k \left\{ \sum_{r=0}^{r_k} \left( \frac{g_k^{(r)}}{G_k} \right) \ln \left( \frac{g_k^{(r)}}{G_k} \right) - \sum_{r=0}^{r_k} \left( \frac{g_k^{(r)}}{G_k} \right) \ln (q_k)^r \right\}, \tag{8.4}$$

$$\frac{S_k}{k_B} = -G_k \left\{ \sum_{r=0}^{r_k} w_k^{(r)} \ln w_k^{(r)} - \sum_{r=0}^{r_k} w_k^{(r)} \ln (q_k)^r \right\}. \tag{8.5}$$

We can now give the entropy $\bar{S}_k$ per state:

$$\bar{S}_k = \frac{S_k}{G_k} = -k_B \sum_{r=0}^{r_k} w_k^{(r)} \ln \left( \frac{w_k^{(r)}}{q_k^r} \right) = -k_B \sum_{r=0}^{r_k} w_k^{(r)} \ln \rho_k^{(r)}. \tag{8.6}$$

In this entropy $\bar{S}_k$ of the quantum gas, besides the familiar probabilities $w_k^{(r)}$ used in computing the mean values, there appear conditional probabilities $\rho_k^{(r)}$:

$$\bar{S}_k = -k_B \sum_{r=0}^{r_k} w_k^{(r)} \ln \rho_k^{(r)} = -k_B \sum_{r=0}^{r_k} \overline{\ln \rho_k^{(r)}} \tag{8.7}$$

$$w_k^{(r)} = \frac{g_k^{(r)}}{G_k} \quad , \quad \rho_k^{(r)} = \frac{w_k^{(r)}}{q_k^{(r)}} \quad , \quad q_k^{(r)} = (q_k)^r \tag{8.8}$$

The $w_k^{(r)}$ are given in (3.12) and the $q_k$ in (5.7). Whereas $w_k^{(r)}$ is the probability of finding in the $|k\rangle$ bundle a state of the energy $\varepsilon_k$ occupied by $r$ quantum particles, $q_k$ denotes the elementary probability of finding in the quantum gas as a whole a $|k\rangle$ bundle of states in which each quantum particle has an energy in the interval of $0 \leq \varepsilon_k \leq E_*$.

Thus, the conditional probabilities $\rho_k^{(r)}$ indicate the probability of finding in the $|k\rangle$ bundle of a quantum system a state of the energy $\varepsilon_k$ with $r$ quantum particles that *at the same time* satisfies the condition $\varepsilon_k \leq E_*$ (see also [24]). Because $\overline{S}_k \geq 0$, it follows that in (8.6) with (8.8) always $w_k^{(r)} \leq q_k^{(r)} = (1 - \varepsilon_k / E_*)^r$. The greater the energy $\varepsilon_k \leq E_*$ of the gas particles, the smaller will be the probability of finding states with many $r$ particles.

Now we can show that we obtain the same grand canonical formulas of the thermodynamics of the ideal quantum gas, computed in chapter 4, if we maximize this entropy $\overline{S}_k$ per state, based on the variation of the probabilities $w_k^{(r)}$,

$$\overline{S}_k = \frac{S_k}{G_k} = -k_B \sum_{r=0}^{r_k} w_k^{(r)} \ln \rho_k^{(r)} = -\overline{\ln \rho_k^{(r)}} \Rightarrow \max , \tag{8.9}$$

with $\rho_k^{(r)}$ from (8.8) and $q_k$ from (5.7). The side conditions of the variation $\delta w_k^{(r)}$ now read

$$\sum_{r=0}^{r_k} r w_k^{(r)} = \left[ \sum_{r=0}^{r_k} \frac{r g_k^{(r)}}{G_k} \right] = \frac{N_k}{G_k} , \tag{8.10}$$

$$\sum_{r=0}^{r_k} r w_k^{(r)} \varepsilon_k = \left[ \sum_{r=0}^{r_k} \frac{r g_k^{(r)} \varepsilon_k}{G_k} \right] = \frac{E_k}{G_k} , \tag{8.11}$$

$$\sum_{r=0}^{r_k} w_k^{(r)} = \left[ \sum_{r=0}^{r_k} \frac{g_k^{(r)}}{G_k} \right] = 1 . \tag{8.12}$$

Here, we have again used the definitions (3.2) - (3.4) for the mean occupation number $N_k$, the mean energy $E_k$ and the total number of the states $G_k$. After execution of the variation, there result the thermodynamic relationships for the grand canonical potential $\Phi$ and the thermodynamic functions $P$, $E$, $N$ and $S$ (with $S = \sum_k S_k$, $S_k = G_k \overline{S}_k$) that were already derived in another way in chapter 4.

## 9 Conclusions

The hypothesis (1.2) that Planck's energy $E_* = M_* c^2$ is a natural limit for the size of quantum leaps between the energy eigenstates of a quantum system has been integrated into the thermodynamics of ideal quantum gases. If quantum leaps between the states of a quantum system may not exceed some limit size energetically, the number of possible microstates of this system is restricted. If we take this into account in Boltzmann's principle $S = k_B \ln \Omega$, there result new thermodynamic distribution formulas (6.9) and (6.20) with a parameter $q_k = (1 - \varepsilon_k/E_*)$. In passing to the limit $E_* \to \infty$, these distribution formulas transform into the well-known Fermi and Bose distributions.

If we look at the structure of the associated new fluctuation formulas (6.28) and (6.34), we find the interesting additional term $\varepsilon_k \cdot \bar{\varepsilon}_k / E_*$ there. Due to this term, the quantum-mechanical particle or wave character of the gas in the bundle of states $|k\rangle$ can be changed by merely selecting a suitable size of the 1-particle-energy $\varepsilon_k$ or the mean particle energy $\bar{\varepsilon}_k$, respectively, referred to the Planck energy $E_*$.

The new distribution formulas with Planck's limit energy $E_*$ have then been applied to ideal quantum gases of the Fermi and Bose type.

If the gas particles with rest masses $m_0$ are moved in the non-relativistic mode, there appear, in the thermal and caloric equations of state of the quantum gas, quantum corrections that are effected by the factor $q_0 = (1 - m_0/M_*)$. As the particle masses are, usually very small compared to the Planck mass $M_*$, then is $q_0 \approx 1$, and the changes in the equations of state (7.15) of the classical limit case of ideal quantum gases are negligibly small. Remarkably new statements, however, are obtained in the low-temperature case $T \to 0$. In the Fermi gas, the mean occupation number $\bar{n}_0$ of the ground state $\varepsilon = 0$ at $T = 0$ has the form $\bar{n}_0 = 1/(1 - (m_0/M_*))$. Thus it differs from exactly $\bar{n}_0 = 1$ which we would expect according to Pauli's principle. If we increase the rest mass $m_0$ of the particles up to Planck's mass $M_*$, then the mean occupation number $\bar{n}_0$ of the ground level can increase to any size: With extremely massive fermions, the Fermi gas takes on bosonic properties. In the Bose gas, likewise, the mean occupation number of the ground state $\varepsilon = 0$ at $T = 0$ has the form $\bar{n}_0 = M_*/m_0 = N_0^{max}$. It indicates that, in the ground level of the Bose gas, i.e., in a Bose-Einstein condensate, there can exist maximally $\bar{n}_0 \leq N_0^{max}$ particles, rather than any number of particles $\bar{n}_0 \to \infty$ as otherwise common in the Bose case. If here we increase the rest mass $m_0$ of the particles up to Planck's mass $M_*$, the mean occupation number of the ground state approaches the value $\bar{n}_0 = 1$: With extremely massive bosons, the Bose gas takes on fermionic properties.

These statements are in accordance with the following properties of the new fermionic and bosonic distributions (7.1) – (7.3): If the rest masses $m_0$ of the gas particles increase up to Planck's mass $M_*$, $m_0 \to M_*$, $q_0 = (1 - (m_0/M_*)) \to 0$, both distributions change into the Boltzmann distribution of an ideal gas with purely classical thermodynamic behaviour (equation of state $PV = N k_B T$). For such a Planck mass gas, the gas degeneration does no longer take place: In the Fermi gas the Fermi momentum vanishes, $p_F = 0$ (cf. (7.17)), and in the Bose gas the critical Einstein temperature disappears, $T_E = 0$ (cf. (7.19)). We note: Due to passing to the limit $m_0 \to M_*$ ($q_0 \to 0$), all quantum effects in ideal quantum gases are suppressed; the ideal quantum gases turn into classical gases of the Boltzmann type.

If the gas particles move in the ultrarelativistic mode, $cp \gg \varepsilon_0 = m_0 c^2$, one can apply the energy-momentum relation $\varepsilon = cp$ in the distributions (7.1) – (7.3). For the case $k_B T \ll E_*$, the known properties of the ideal ultrarelativistic quantum gases are maintained.

New statements, however, result for the case $k_B T \gg E_*$, i.e., for $T \to \infty$. These are:

The mean energy $\bar{\varepsilon}$ of the energy levels $\hbar \omega$ is limited by $\bar{\varepsilon} \leq E_*$.

With increasing temperatures, the maximum of the spectral energy distribution $(\hbar \omega)_{max}$ shifts, relative to Wien's known displacement law, towards higher frequencies – slower in the case of fermions and faster in the case of bosons.

The internal energy $U$, the particle number $N$ and the entropy $S$ have, for $T \to \infty$, finite limit values ($N_* = V/V_*$):

$$U \xrightarrow[T \to \infty]{} N_* E_* \quad , \quad N \xrightarrow[T \to \infty]{} N_* \quad , \quad S \xrightarrow[T \to \infty]{} N_* k_B \quad . \tag{9.1}$$

For $T \to \infty$, the equation of state has the simple form

$$PV \xrightarrow[T \to \infty]{} N_* k_B T \quad . \tag{9.2}$$

Especially the property of a constant energy density $u = U/V$ of the photon or neutrino gas at extremely high temperatures following from (9.1),

$$u(T) \xrightarrow[T \to \infty]{} u_* \quad , \tag{9.3}$$

with $u_* = E_*/V_*$ (Planck´s energy density) has a decisive effect on the behaviour of the early Friedmann universe: Cosmic expansion starts from a space region having the linear size of one Planck length $L_*$, i.e., a big bang does not exist in that case [18].

Investigating ideal quantum gases, we came to some new physical statements, provided that the hypothesis (1.2) of the existence of an energetically maximum possible quantum leap is implemented in nature. Perhaps one or the other of the theoretical results obtained herein can be experimentally verified in the future. It would then furnish first clues for quantum gravity effects.

**Appendix**

In the diagrams Fig. 5 – Fig. 11 it can be observed that significant deviations from the todays valid statistics of ultrarelativistic ideal quantum gases should occur only at temperatures $\theta > 0.1$, i.e., about at $k_B T > E_*/10$.

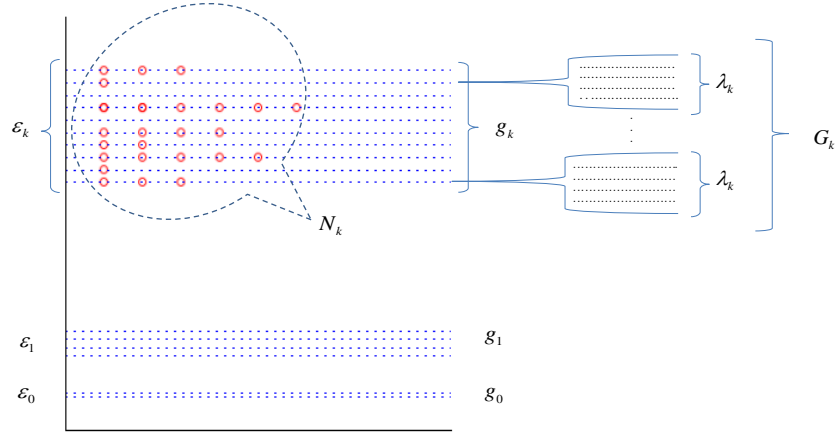

Fig. 1: Bundle of states $|k\rangle$ with $g_k$ states of energy $\varepsilon_k$ and $N_k$ quanta in them. Each of the $g_k$ states is supposed to have, in addition, a $\lambda_k$ - fold degeneration caused by a fluctuating microstructure of the spacetime manifold.

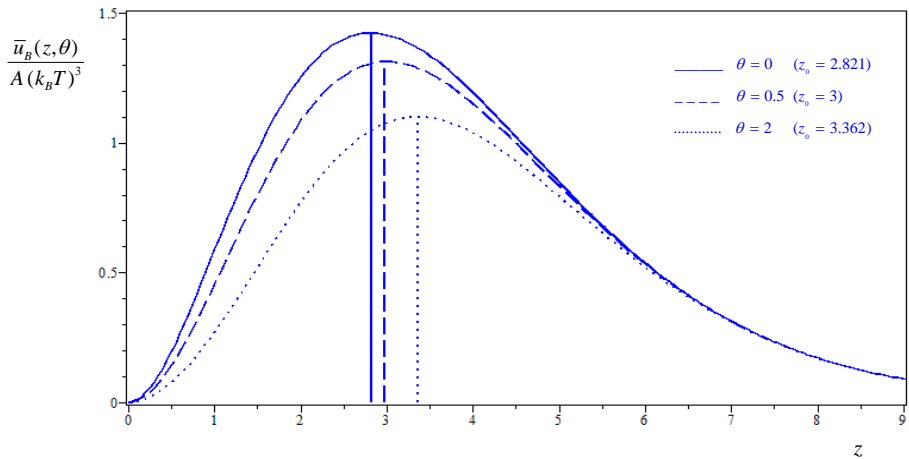

Fig. 2: Spectral energy density $\bar{u}_B(z,\theta)$ of ultra-relativistic bosons (photons, …) as a function of their oscillation frequency $z = \hbar\omega/k_B T$ and the temperature parameter $\theta = k_B T/E_*$. The curve with the parameter $\theta = 0$ corresponds to the hitherto valid quantum statistics ($A = g_s/2\pi^2 \hbar^2 c^3$).

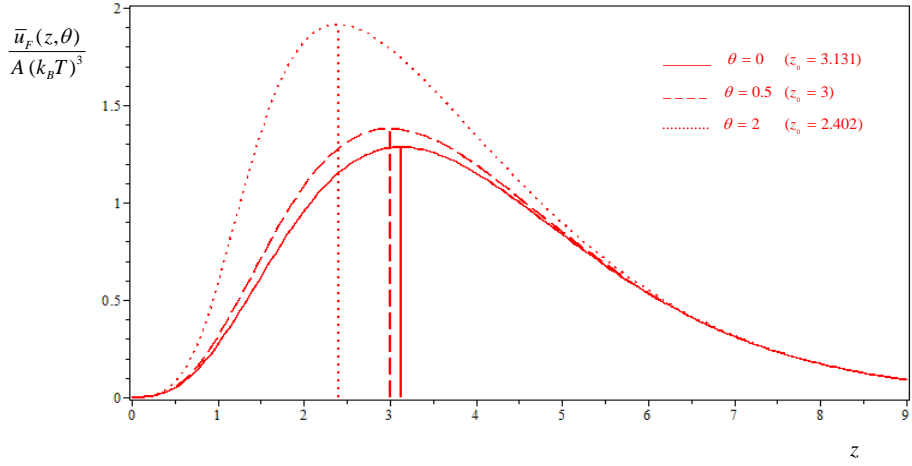

Fig. 3: Spectral energy density $\bar{u}_F(z,\theta)$ of ultra-relativistic fermions (neutrinos, …) as a function of their oscillation frequency $z = \hbar\omega/k_B T$ and the temperature parameter $\theta = k_B T/E_*$. The curve with the parameter $\theta = 0$ corresponds to the hitherto valid quantum statistics ($A = g_s/2\pi^2\hbar^2 c^3$).

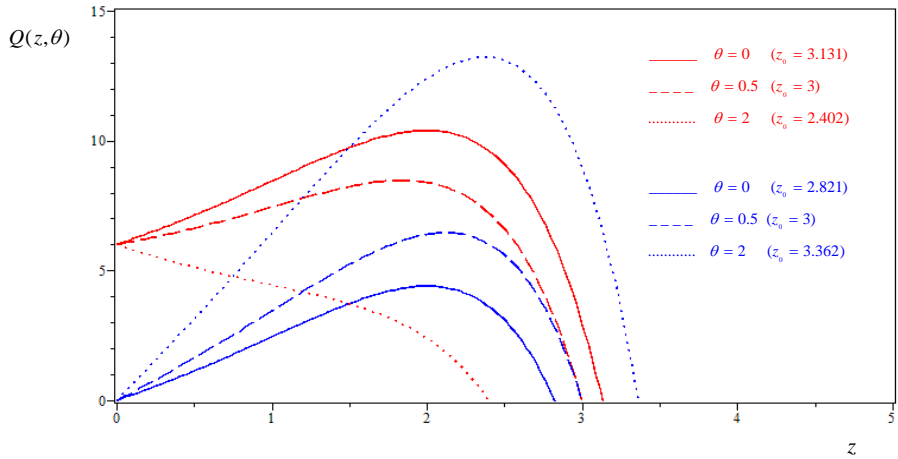

Fig. 4: The zeros $z_0(\theta)$ of the function $Q(z,\theta)$ represent Wien's displacement law for ultrarelativistic quantum gases in the form $(\hbar\omega)_{max} = z_0 \cdot k_B T$. The hitherto valid quantum statistics is characterized by the value $z_0 = z_0(\theta = 0)$. Blue curves describe bosons, red ones fermions.

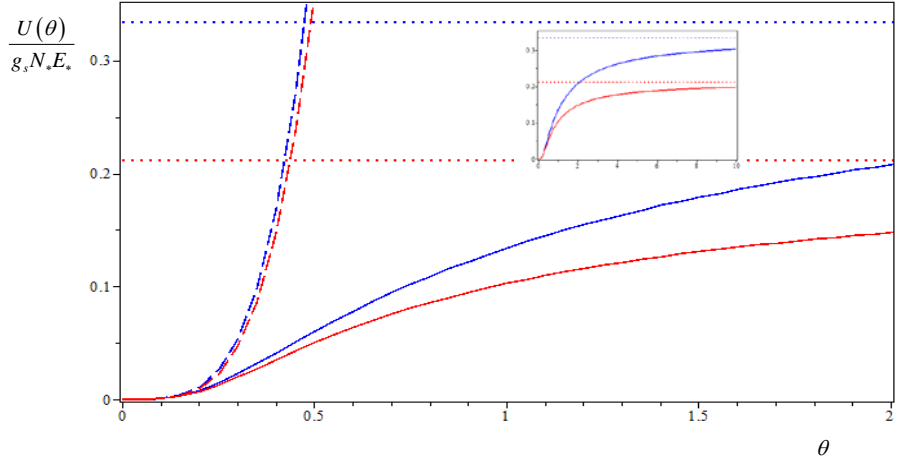

Fig. 5: The internal energy $U(\theta)$ of ultra-relativistic quantum gases as a function of the temperature parameter $\theta = k_B T / E_*$. Blue curves describe bosons, red ones fermions. For extremely high temperatures, both curves have finite limit values: For bosons there results the value $U/(g_s N_* E_*) \xrightarrow[(\theta \to \infty)]{} 1/3$, for fermions $U/(g_s N_* E_*) \xrightarrow[(\theta \to \infty)]{} 8\ln(2) - 16/3 \approx 0.212$. The broken curves indicate the behaviour according to the hitherto valid quantum statistics.

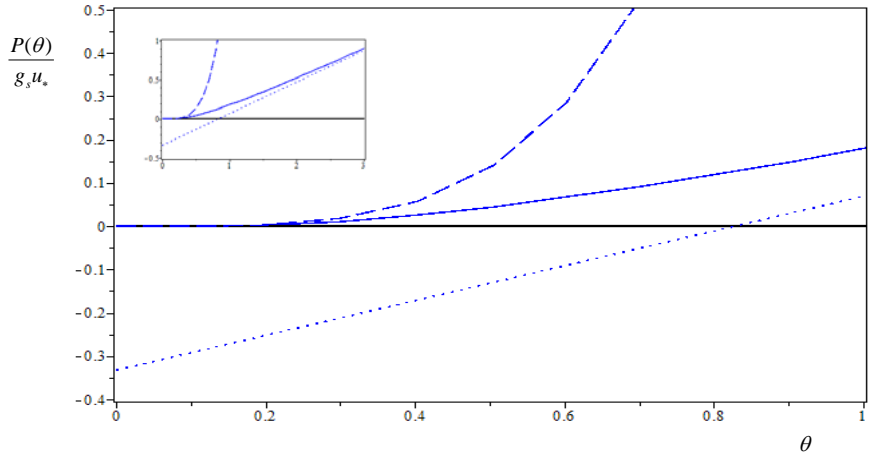

Fig. 6: The pressure $P(\theta)$ of ultra-relativistic Bose gases as a function of the temperature parameter $\theta = k_B T / E_*$. For extremely high temperatures, the curve has an asymptote: $P(\theta)/(g_s u_*) \xrightarrow[(\theta \to \infty)]{} C_\infty^B \cdot \theta - 1/3$ (with $C_\infty^B = 0.395$). The broken curve indicates the behaviour according to the hitherto valid quantum statistics.

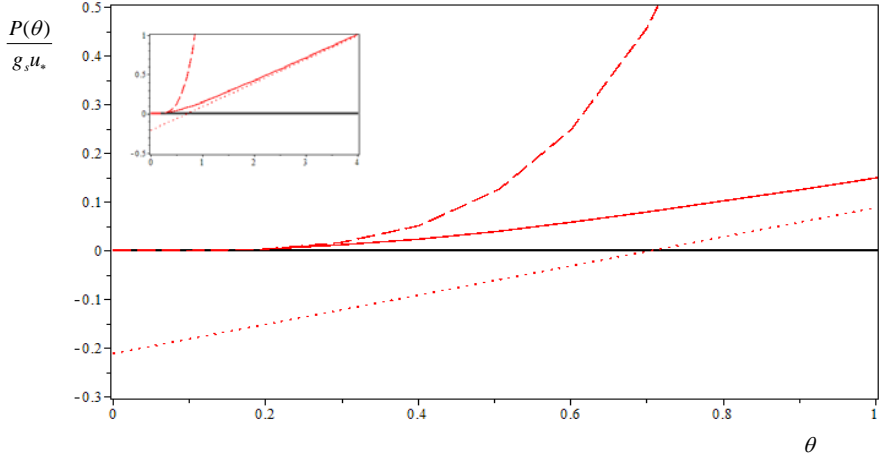

Fig. 7: The pressure $P(\theta)$ of ultra-relativistic Fermi gases as a function of the temperature parameter $\theta = k_B T/E_*$. For extremely high temperatures, the curve has an asymptote: $P(\theta)/(g_s u_*) \xrightarrow[(\theta \to \infty)]{} C_\infty^F \cdot \theta - 0.212$ (with $C_\infty^F = 0.300$). The broken curve indicates the behaviour according to the hitherto valid quantum statistics.

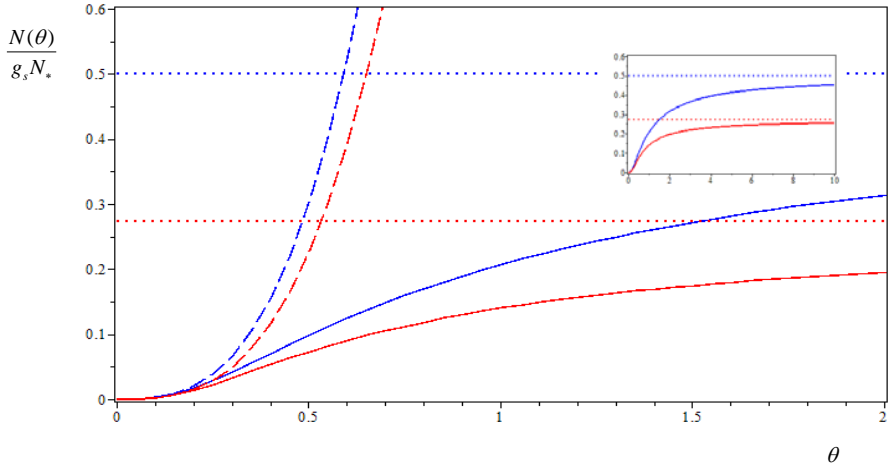

Fig. 8: The total number of particles $N_*(\theta)$ of ultra-relativistic quantum gases as a function of the temperature parameter $\theta = k_B T/E_*$. Blue curves describe bosons, red ones fermions. For extremely high temperatures, both curves have finite limit values: For bosons there results the value $N(\theta)/(g_s N_*) \xrightarrow[(\theta \to \infty)]{} 1/2$, and for fermions result the value $N(\theta)/(g_s N_*) \xrightarrow[(\theta \to \infty)]{} 4\ln(2) - 5/2 \approx 0.273$. The broken curves indicate the behaviour according to the hitherto valid quantum statistics.

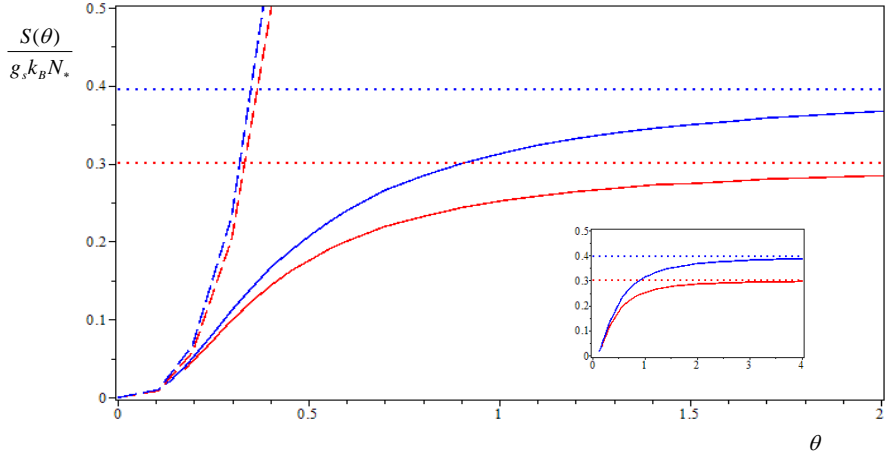

Fig. 9: The entropy $S(\theta)$ of ultra-relativistic quantum gases as a function of the temperature parameter $\theta = k_B T/E_*$. Blue curves describe bosons, red ones fermions. For extremely high temperatures, both curves have finite limit values: For bosons there results the value $S/(g_s k_B N_*) \xrightarrow[(\theta \to \infty)]{} C_\infty^B \approx 0.395$, for fermions $S/(g_s k_B N_*) \xrightarrow[(\theta \to \infty)]{} C_\infty^F \approx 0.300$. The broken curves indicate the behaviour of the entropy according the currently valid quantum statistics.

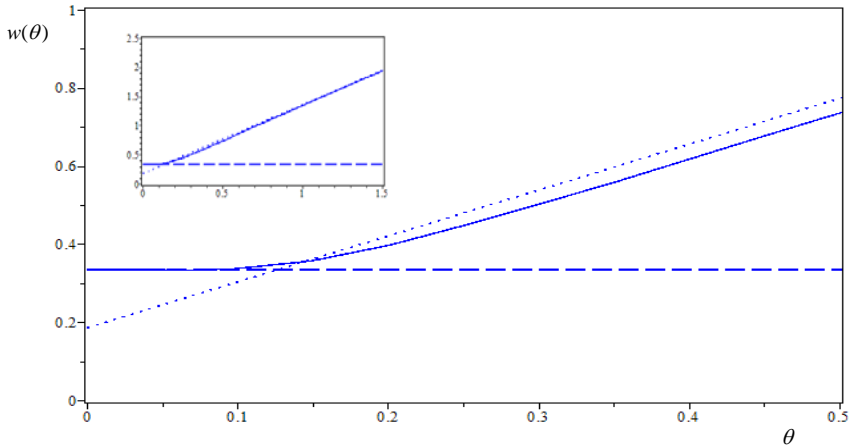

Fig. 10: The ratio $w(\theta) = P(\theta) \cdot V/U(\theta)$ of an ultra-relativistic boson gas as a function of the temperature parameter $\theta = k_B T/E_*$. For extremely high temperatures, the temperature behaviour of $w(\theta)$ shows a linear rise with the asymptote $w(\theta) \xrightarrow[(\theta \to \infty)]{} 3C_\infty^B \cdot \theta + (3C_\infty^B - 1)$. The broken horizontal straight line indicates the constant value $w(\theta) = 1/3$ as computed according to the hitherto valid quantum statistics.

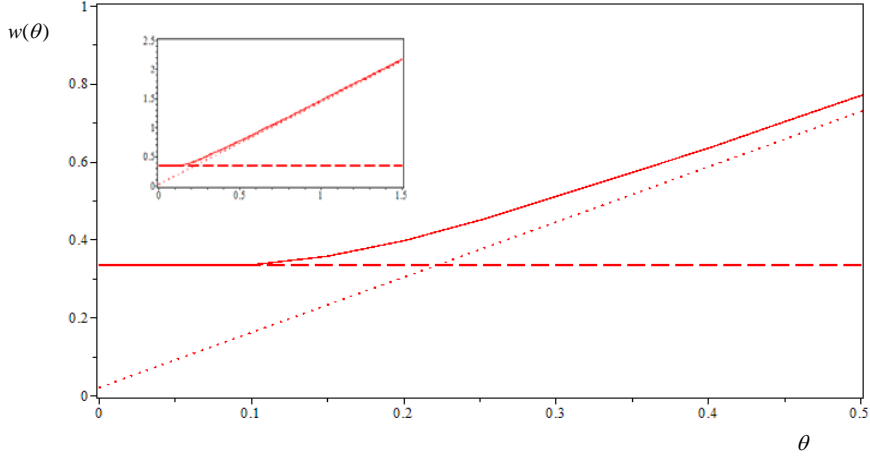

Fig. 11: The ratio $w(\theta) = P(\theta) \cdot V / U(\theta)$ of an ultra-relativistic fermion gas as a function of the temperature parameter $\theta = k_B T / E_*$. For extremely high temperatures, the temperature behaviour of the function $w(\theta)$ shows a linear rise with the asymptote $w(\theta) \xrightarrow[(\theta \to \infty)]{} 4.72 \cdot C_\infty^F \cdot \theta + (3.40 \cdot C_\infty^F - 1)$. The broken horizontal straight line indicates the constant value $w(\theta) = 1/3$ as computed according to the quantum statistics currently valid.

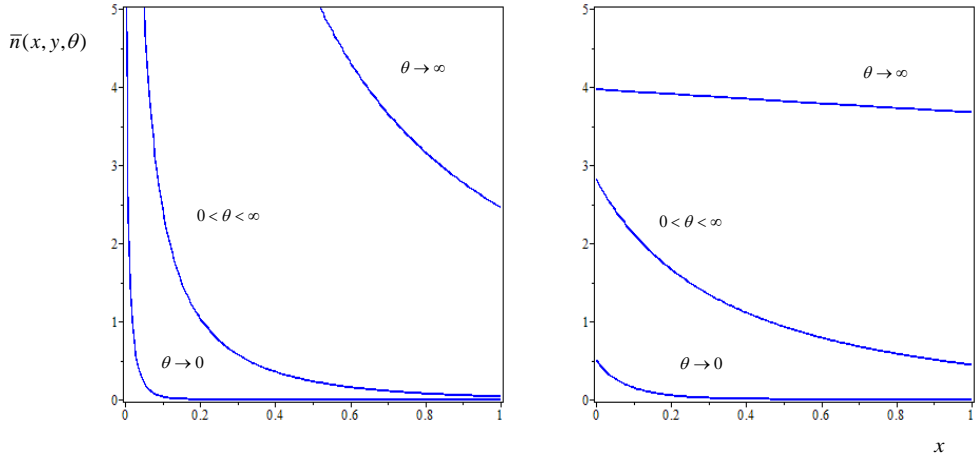

Fig. 12: Distribution of the mean occupation number $\bar{n}(x, y, \theta)$ in a non-relativistic bosonic ideal quantum gas as a function of particle energy $x = \varepsilon / E_*$, with the parameters chemical potential $y = \mu / E_* = const. = -0.10$ and temperature $\theta = k_B T / E_*$. At a rest energy ratio $x_0 = \varepsilon_0 / E_* = m_0 / M_* = 0.01 \ll 1$, the curves in the left diagram correspond to the statements of the hitherto valid quantum statistics with $\bar{n}(x \to 0) \to \infty$. If the mass ratio is changed into $x_0 = 0.25 \sim 1$, though, all curves, according to the new statistics, start at $x = 0$ with a finite value $\bar{n} < \infty$ for the occupation number, as shown in the right diagram (cf. chapter 7.1.3).

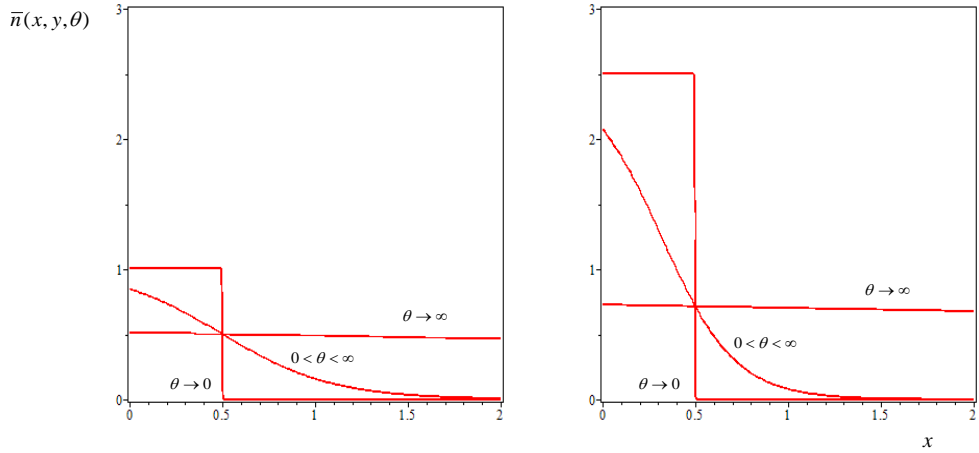

Fig. 13: Distribution of the mean occupation number $\bar{n}(x, y, \theta)$ in a non-relativistic fermionic ideal quantum gas as a function of the particle energy $x = \varepsilon/E_*$ with the parameters chemical potential $y = \mu/E_* = const. = 0.50$ and temperature $\theta = k_B T/E_*$. At a rest energy ratio $x_0 = \varepsilon_0/E_* = m_0/M_* = 0.01 \ll 1$, the curves in the left diagram correspond to the statements of the hitherto valid quantum statistics with $\bar{n}(x=0) \simeq 1$. If the mass ratio is changed into $x_0 = 0.60 \sim 1$, though, all curves, according to the new statistics, start at $x = 0$ with increasing values $\bar{n} > 1$, as shown in the right diagram (cf. chapter 7.1.3).